\documentclass[10pt,journal,compsoc]{IEEEtran}
\newif\ifpeerreview

\peerreviewfalse

\usepackage{lipsum} 

\usepackage[nocompress]{cite}
\usepackage{url}
\usepackage{amsmath,amssymb,graphicx}

\usepackage[switch]{lineno}
\usepackage{color}
\usepackage{ulem}
\usepackage{float}
\usepackage{siunitx}
\usepackage[colorlinks,linkcolor=black,anchorcolor=black,citecolor=black,urlcolor=black]{hyperref}

\DeclareMathOperator{\di}{d\!}

\newcommand{\paperID}{19}

\title{Non-line-of-sight Imaging via \\ Neural Transient Fields}

\author{Siyuan~Shen$^\dag$, Zi~Wang$^\dag$, Ping~Liu, Zhengqing~Pan, Ruiqian~Li, Tian~Gao, \\
        Shiying~Li,
        and~Jingyi~Yu,~\IEEEmembership{Fellow,~IEEE} 

\IEEEcompsocitemizethanks{ 
\IEEEcompsocthanksitem
$^\dag$ indicates equal contribution.\\
\IEEEcompsocthanksitem S. Shen is with the School of Information Science and Technology, ShanghaiTech University, Shanghai 201210, China, and also with DGene, Inc., Shanghai, China. E-mail: {shensy}@shanghaitech.edu.cn.\protect\\
\IEEEcompsocthanksitem Z. Wang is with the Shanghai Institute of Technical Physics, Chinese Academy of Sciences, Shanghai 200083, China, and the School of Information Science and Technology, ShanghaiTech University, Shanghai 201210, China, and also with the University of Chinese Academy of Sciences, Beijing 100049, China. E-mail: {wangzi}@shanghaitech.edu.cn.\protect\\
\IEEEcompsocthanksitem 
P. Liu and R. Li are with the School of Information Science and Technology, ShanghaiTech University, Shanghai 201210, China. E-mail: {\{liuping, lirq1\}}@shanghaitech.edu.cn.\protect\\
\IEEEcompsocthanksitem Z. Pan is with the School of Information Science and Technology, ShanghaiTech University, Shanghai 201210, China, and the Shanghai Institute of Microsystem and Information Technology, Chinese Academy of Sciences, Shanghai 200050, China, and also with the University of Chinese Academy of Sciences, Beijing 100049, China. E-mail: {panzhq}@shanghaitech.edu.cn.\protect\\
\IEEEcompsocthanksitem 
T. Gao is with the School of Physical Science and Technology, ShanghaiTech University, Shanghai 201210, China. E-mail: {gaotian}@shanghaitech.edu.cn.\protect\\
\IEEEcompsocthanksitem 
S. Li, J. Yu are with Shanghai Engineering Research Center of Intelligent Vision and Imaging, and also with the School of Information Science and Technology, ShanghaiTech University, Shanghai 201210, China. E-mail: {\{lishy1, yujingyi\}}@shanghaitech.edu.cn.\protect\\
(Corresponding authors: Shiying Li; Jingyi Yu).
}
}
\begin{document}

\IEEEtitleabstractindextext{%
\begin{abstract}
We present a neural modeling framework for non-line-of-sight (NLOS) imaging. Previous solutions have sought to explicitly recover the 3D geometry (e.g., as point clouds) or voxel density (e.g., within a pre-defined volume) of the hidden scene. In contrast, inspired by the recent Neural Radiance Field (NeRF) approach, we use a multi-layer perceptron (MLP) to represent the neural transient field or NeTF. However, NeTF measures the transient over spherical wavefronts rather than the radiance along lines. We therefore formulate a spherical volume NeTF reconstruction pipeline, applicable to both confocal and non-confocal setups. Compared with NeRF, NeTF samples a much sparser set of viewpoints (scanning spots) and the sampling is highly uneven. We thus introduce a Monte Carlo technique to improve the robustness in the reconstruction. Experiments on synthetic and real datasets demonstrate NeTF achieves state-of-the-art performance and can provide reliable reconstructions even under semi-occlusions and on non-Lambertian materials. 

\end{abstract}

}

\maketitle
\thispagestyle{empty}

\IEEEraisesectionheading{
  \section{Introduction}\label{sec:introduction}
}

\IEEEPARstart{N}{on}-line-of-sight (NLOS) imaging employs time-resolved measurements for recovering hidden scenes beyond the direct line of sight from a sensor \cite{2017Jarabo, 2020Faccio}. Applications are numerous, ranging from remote sensing to autonomous driving and to rescue missions in hazardous environments. Most existing NLOS setups orient an ultra-fast pulsed laser beam towards a relay wall in the line of sight where the wall diffuses the laser into spherical wavefronts towards the hidden scene. As the wavefront hits the scene and bounces back onto the wall, a time-of-flight (ToF) detector with picosecond resolution (such as the streak camera \cite{2012Velten, 2012Gupta, 2013Velten} or recently the more affordable single-photon avalanche diodes (SPADs) \cite{2015Mauro, 2017Tsai, 2018Xu, 2019FK, 2019Xin, 2018LCT}) can be used to record the arrival time and the number of the returning photons \cite{2015Gariepy, 2016Shin, 2017Matthew}. SPAD sensors in a time correlated single photon counting (TCSPC) mode can thus produce transients, of which a single pixel corresponds to a specific pair of illumination and detection spots on the wall and a histogram of the photon counts versus time bins. 

\begin{figure}[!t]
\centering
\includegraphics[width=0.5\textwidth]{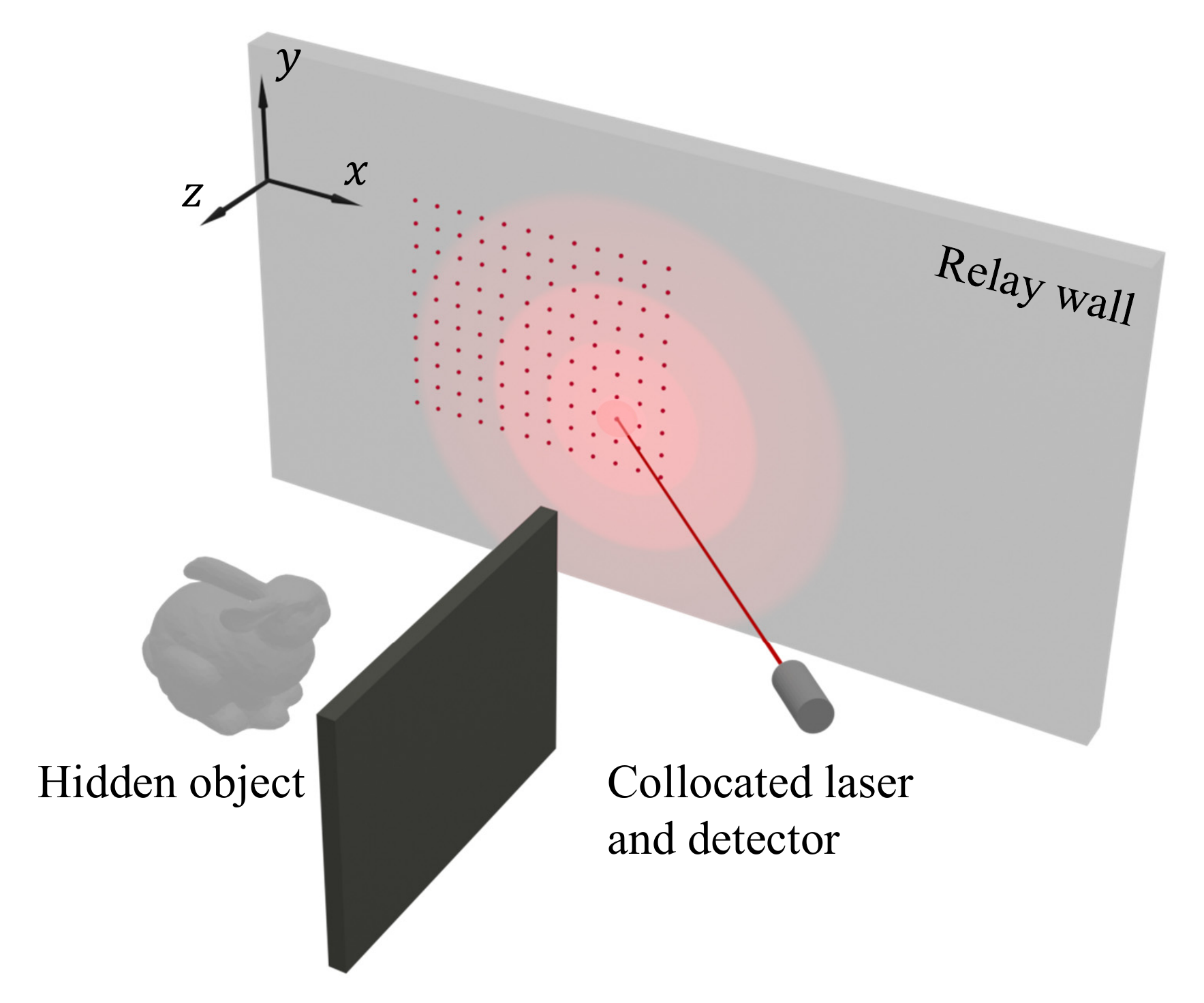}
\caption{A typical confocal NLOS imaging system aims a laser towards a diffuse wall that serves as a virtual reflector. The hidden scene is indirectly illuminated as
spherical waves that intersect with the scene and are reflected back onto the wall. A SPAD sensor measures at different spots on the wall to form transient images.}
\label{SystemSetting}
\end{figure}

The measured transients contain rich geometric information of a hidden scene, potentially usable for scene recovery. In reality, the process corresponds to a typical inverse imaging problem that generally incurs high computational cost, especially because the transients are high dimensional signals. To make the problem tractable, the pioneering back-projection (BP) technique and its variations assume smooth objects so that scene recovery can be modeled as deconvolution \cite{2012Velten, 2012Gupta, 2015Mauro, 2017Victor}. Alternatively, the light-cone transform (LCT) based methods collocate the illumination and sensing spots on the relay wall so that the forward imaging model can be simplified as 3D convolution \cite{2018LCT, 2019FK, 2020DLCT, 2020Mariko} where advanced signal processing techniques such as Wiener filters \cite{2018LCT, 2019Ahn, 2020DLCT} can further reduce noise. Assuming the scene is near diffuse, analysis-by-synthesis algorithms can improve reconstruction \cite{2019Tsai, 2020ToG}. The seminal work of the Fermat path based approaches \cite{2017Tsai, 2019Xin} can handle highly specular objects by simultaneously recovering the position and normal of Fermat points on the surface.  

\begin{figure*}[htbp]
\centering
\includegraphics[width=0.87\textwidth]{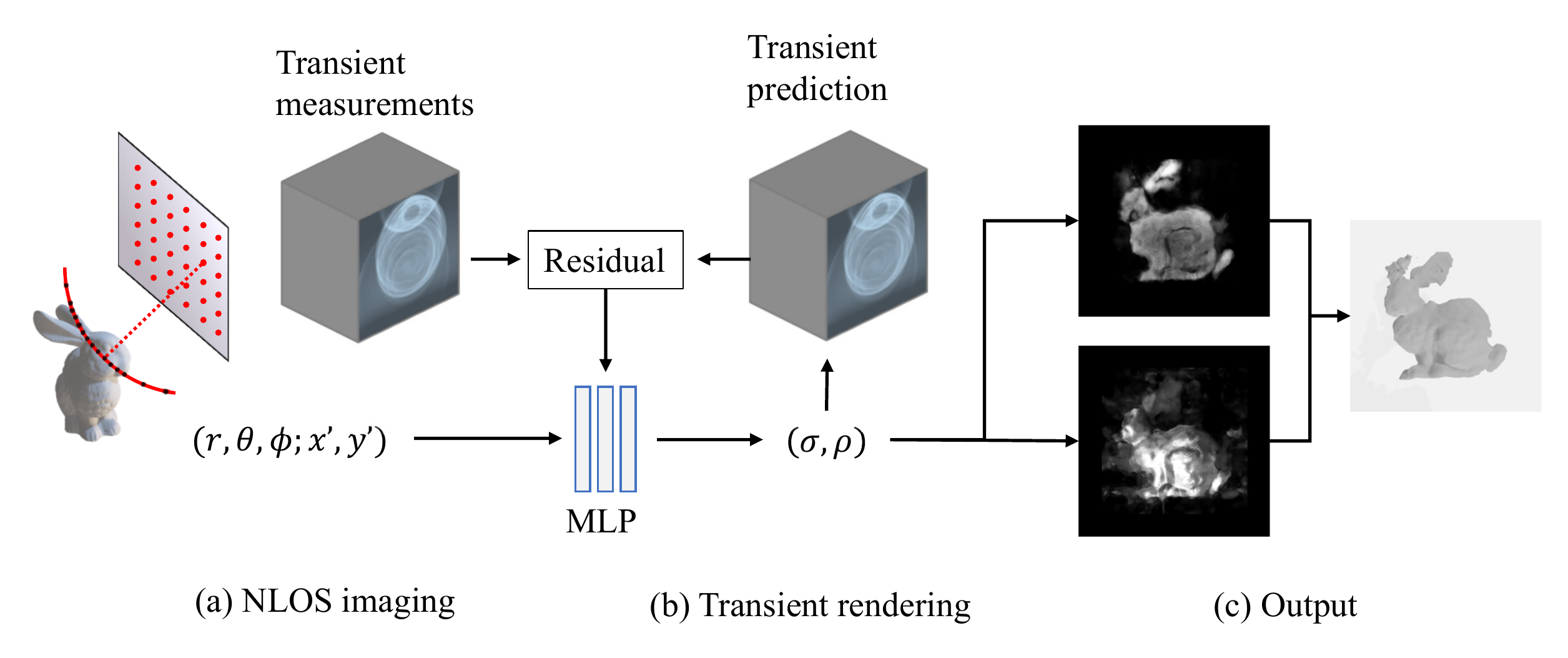}
\caption{Our neural transient field (NeTF) reconstruction pipeline. We parameterize every point on a spherical wavefront in terms of the origin on the wall, and the direction and radius of its corresponding spherical coordinates. We set out to recover the transient field under this parameterization via a multi-layer perception under spherical volume rendering.}
\label{Pipeline}
\end{figure*}

We present a novel volumetric NLOS imaging framework by modeling the transient field via deep networks. Our Neural Transient Field (NeTF) technique is inspired by the recent neural radiance field (NeRF) that conducts 3D reconstruction and view synthesis from a set of input images. Different from existing multi-view stereo (MVS) techniques, NeRF assumes a volume rendering model and sets out to use multi-layer perception (MLP) to recover per-voxel scene density and per-direction color. We observe that NLOS resembles MVS in that each scanning point on the wall resembles a virtual camera and therefore a similar deep learning technique may be potentially used for scene recovery. Different from NeRF, though, NeTF measures the transient over spherical wavefronts rather than the radiance along lines. We therefore first formulate volumetric transient fields under the spherical coordinate and devise an MLP that trains on the measurements to predict per-voxel density and view-dependent albedo. Our NeTF formulation is applicable to both confocal and non-confocal setups. 

\begin{figure}[!h]
\centering
\includegraphics[width=0.5\textwidth]{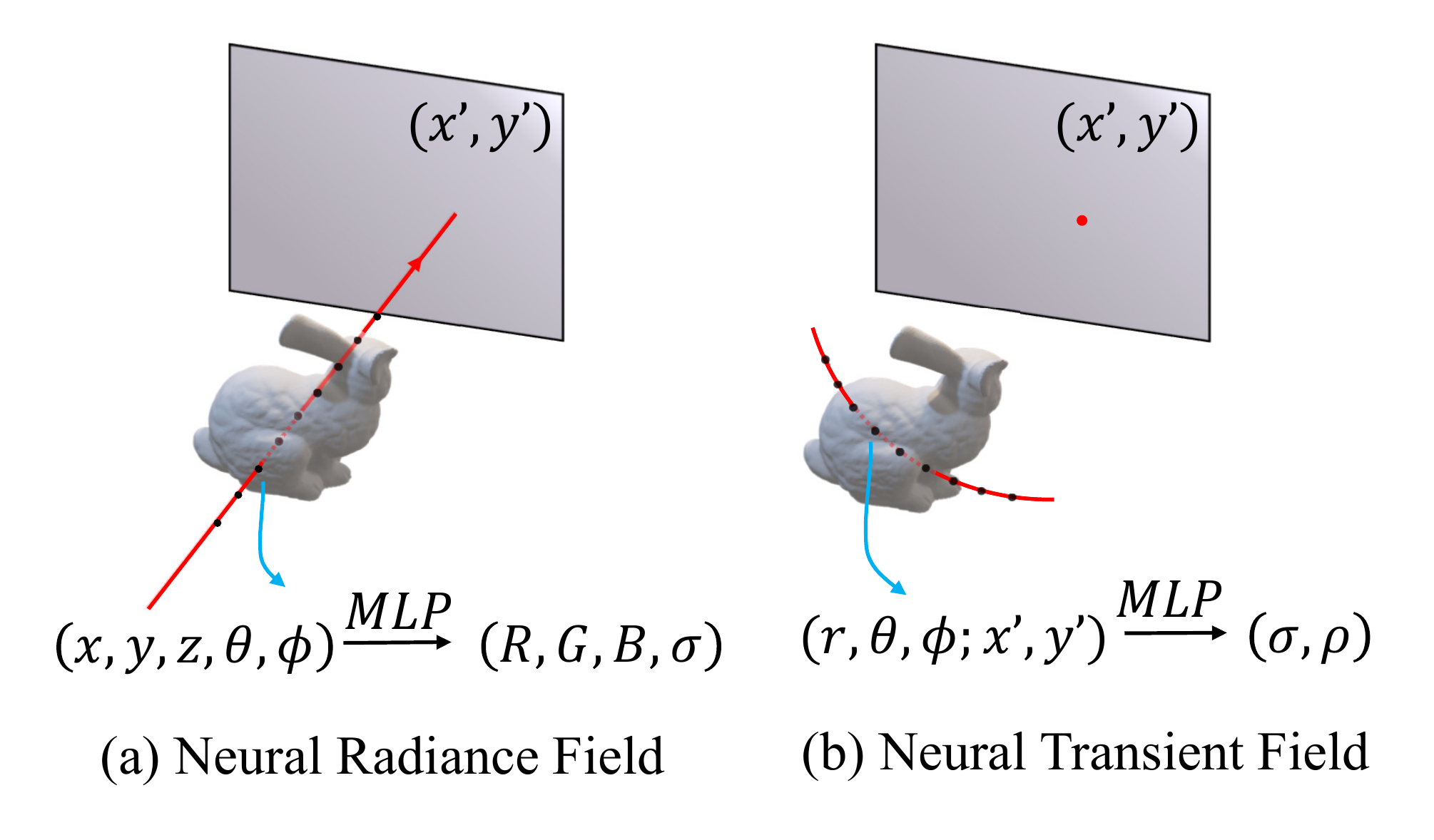}
\caption{NeRF vs. NeTF. In NeRF, volume density is accumulated along every line (ray) whereas in NeTF it is accumulated on a spherical wavefront. }
\label{NeRFvsNLOS}
\end{figure}

Compared with NeRF, NeTF captures a much sparser set of viewpoints (scanning spots) and the distribution of scene points on the spherical wavefronts can be highly uneven. We therefore develop a Markov chain Monte Carlo (MCMC) technique based on importance sampling for matching the actual scene distribution. We conduct comprehensive experiments on existing synthetic and real datasets. We demonstrate that NeTF achieves state-of-the-art reconstruction quality under both confocal and non-confocal settings. In particular, the trained MLP provides a continuous 5D representation of the hidden scene without requiring digitizing the NLOS volume or optimizing surface parameters. The 5D representation with directional encoding can handle view-dependent albedo of non-Lambertian surface reflectance and strong self-occlusions, under both confocal and non-confocal setups. All our codes and data are available at \url{https://github.com/zeromakerplus/NeTF_public}.


\section{Related Work}

As an emerging computational imaging technique, NLOS imaging has found broad applications in computer vision and computer graphics, ranging from recovering 3D shape of hidden objects \cite{2012Velten, 2015Mauro, 2017Tsai, 2018LCT, 2019FK, 2019Liu, 2019Heide, 2020DLCT} to tracking hidden moving objects \cite{2012Velten, 2018Neural, 2019Maeda, 2020Mariko}. Existing solutions employ time-resolved optical detectors such as streak cameras \cite{2012Velten, 2012Gupta}, SPADs \cite{2015Mauro, 2015Gariepy} and interferometry \cite{2015Ioannis, 2019Xin} or non-optical acoustic \cite{2019Lindell} and thermal \cite{2019Maeda} sensors to indirectly measure the hidden scene and then apply inverse imaging techniques for recovery. We refer readers to recent surveys \cite{2017Jarabo, 2018Yoann, 2020Faccio} for a comprehensive overview. 

\textbf{Confocal vs. Non-Confocal.} Kirmani et al. \cite{ 2011Kirmani, 2009Kirmani} designed and implemented the first prototype non-confocal NLOS system and derived a linear time-invariant model amenable to multi-path light transport analysis. In reality, varying both the laser beam and the measuring spot yield a high-dimensional transient field analogous to the light field. Many efforts have since been focused on imposing priors and constraints to accelerate data processing. Velten et al. \cite{2012Velten} proposed a back-projection technique with ellipsoidal constraints: the observing point and the laser projection point on the wall correspond to the foci of a set of ellipsoids, each corresponding to a specific transient. The hidden scene can then be reconstructed by intersecting the ellipsoids. To further improve reconstruction quality and speed, subsequent work has applied filtering techniques such as sharpening and thresholding \cite{2012Gupta, 2015Mauro, 2017Victor}. Alternatively, one can directly model the scene using parametric surfaces and then optimize the parameters over the observations \cite{2019Tsai, 2020ToG, 2019Ahn}. Ahn et al. \cite{2019Ahn} model parameter fitting as a linear least-squares problem using a convolutional Gram operator. It is also possible to adopt wave optics for NLOS imaging \cite{2019Reza, 2019Liu, 2019FK, 2020Liu}, by characterizing the problem as specific properties of a temporally evolving wave field in the Fourier domain.  

To reduce data dimensionality, several recent approaches adopt a confocal setting \cite{2018LCT, 2020DLCT} where the laser and the detector (e.g., a SPAD) collocate, e.g., via a beam splitter. Consequently, the ellipsoidal constraints degenerate to be spherical, simplifying the inverse problem with a 3D deconvolution and system calibration. The seminal work of light-cone transform (LCT) \cite{2018LCT} casts the NLOS reconstruction problem as Wiener filtering in the Fourier domain and can achieve a low computational complexity of $O(N^3\log{N})$ for $N^3$ voxels, compared to $O(N^5)$ in the traditional BP methods. Yong et al. \cite{2020DLCT} formulate the albedo and normal recovery based on directional LCT (DLCT) as a vector deconvolution problem. The confocal setting results in an overwhelming contribution of direct light first-bounce off the diffuse wall and subsequently produces useful geometric constraints. The seminal work by Xin et al. \cite{2019Xin} exploits the Fermat flow induced by the transients for estimating surface normals. Lindell et al. \cite{2019FK} adapt an F-K migration in seismology to convert the surface reconstruction problem to a boundary value problem. The F-K migration method enables faster reconstruction and supports planar or nonplanar diffuse walls. 



\textbf{Volume vs. Surface.} Existing NLOS methods can also be categorized in terms of the form of the reconstruction results. Two most adopted forms are volume density and points/surfaces. Methods for recovering the former generally discretize the scene into voxels and compute the density, either using intersections of wavefronts under ellipsoidal \cite{2012Velten, 2012Gupta, 2015Mauro, 2015Gariepy, 2017Victor, 2019Ahn, 2019Manna} and spherical \cite{2018LCT, 2020DLCT} constraints, or via modeling the imaging process as convolution and recovering the volume via specially designed deconvolution filters. Methods for recovering the latter have relied on light transport physics \cite{2019Tsai, 2020ToG} for optimizing the shape and reflectance of the hidden scenes. Such methods are generally mathematically tractable but are computationally expensive particularly because higher order geometry such as the surface normal needs to integrated into the optimization process. Reconstruction results are either sparse as in Fermat \cite{2019Xin} where only discontinuities in the transient were used, or rely heavily on the quality of the basis shape as in \cite{2019Tsai}.

Our approach falls into the category of volume based technique. We are inspired by the recent multi-view reconstruction framework Neural Radiance Field (NeRF) that aims to recover the density and color at every point along every ray, implicitly providing a volumetric reconstruction. NeRF adopts a volume rendering model and sets out to optimize volume density that best matches the observation using a Multi-Layer Perception (MLP). It is also possible to modify NeRF to tackle photometric stereo (PS) problems where the camera is fixed but the lighting conditions vary. We observe the non-confocal NLOS imaging process greatly resembles MVS/PS: fixing the laser beam but measuring the transient at different spots on the wall resembles MVS, whereas fixing the measuring spot but varying the laser beam resembles PS. In fact, the confocal setting is very similar to the NeRF AA setting \cite{bi2020neural} where the lighting and the camera move consistently. We therefore call our reconstruction scheme Neural Transient Field or NeTF.

Both NeTF and NeRF use MLP as an optimizer. However, there are several major differences between NeRF and NeTF. First, the volume rendering model used in NeRF is not directly applicable to NeTF. We therefore derive a novel volumetric image formation model under NLOS. Second, NLOS measures the transient rather than the radiance. In fact, the transient is measured from the sum of returning photons on a wavefront instead of a single ray. We hence formulate a spherical volume reconstruction pipeline. Finally, NeRF generally assumes dense ray samples, whereas the NLOS setting is much more sparse. We thus introduce a Monte Carlo technique to improve the robustness in the reconstruction.

\section{Neural Transient Field}

We recognize that the NLOS reconstruction problem resembles multi-view reconstruction in the line-of-sight (LOS) and adopt a neural reconstruction framework analogous to NeRF \cite{2020NERF}. Each detection spot on the relay wall can be viewed as a virtual \textit{camera}. These cameras capture the transients of the NLOS scene as if viewed from the wall. We adopt the plenoptic radiance field notion of NeRF and represent the NLOS scene as a continuous 5D function of transients, i.e., a plenoptic transient field. We then set out to infer scene density at every point along every spherical wavefront via deep network based optimization. It is important to note that, same as NeRF, our neural transient field (NeTF) representation chooses not to explicitly discretize the scene into volumes. Rather, we use multi-layer perception (MLP) to virtually represent the volume. 



\subsection{Scene Representation}
The NeRF framework \cite{2020NERF} uses the neural radiance field $L(x, y, z, \theta, \phi)$ as scene representation where $(x, y, z)$ corresponds to a point on a ray and $(\theta, \phi)$ the direction of the ray. Its trained network outputs both the density $\sigma$ at every position $(x, y, z)$  and the (view-dependent) color $\boldsymbol{c} = (r, g, b)$ along direction $(\theta, \phi)$. The density can be further used for scene reconstruction and the color for image-based rendering. In our case, NeTF, instead of sampling on a single camera ray, samples a hemisphere of rays as light propagates as a spherical wave from the relay wall towards the hidden scene. We hence adopt a continuous 5D function of transients $L_{\text{NLOS}}$ under the spherical coordinates as:

\begin{equation}
L_{\text{NLOS}}(x', y', r, \theta, \phi) \to (\sigma, \rho)
\end{equation}

\noindent where $P(x', y')$ is a detection spot on the wall that serves as the origin of the hemisphere. $Q(r, \theta, \phi)$ is a scene point parameterized using the spherical coordinate $(r, \theta, \phi)$ w.r.t. $P(x', y')$. Similar to NeRF, though, we set out to design a fully connected neural network, i.e., an MLP, to estimate $L_{\text{NLOS}}$. Different from NeRF, $L_{\text{NLOS}}$ in NeTF outputs a volume density $\sigma$ and a surface reflectance (albedo $\rho$) rather than color along the direction $(\theta, \phi)$. 

Recall that NLOS needs to scan different spots on the relay wall, resulting in inconsistent spherical coordinates and casting challenges in network training and inference. We thus first transform the spherical coordinates $(x',y', r, \theta,\phi)$ to their corresponding Cartesian coordinates, i.e.,  $(x, y, z, \theta,\phi)$ as 

\begin{equation}
R:
 \left\{
             \begin{array}{lr}
             x=r \sin\theta \cos\phi + x' &  \\
             y=r \sin\theta \sin\phi + y' & \\
             z=r \cos\theta &  
             \end{array}
\right.
\label{transform}
\end{equation}

The transform $R$ ensures that the position of a 3D voxel is consistent when we scan over different detection spots. All subsequent training under MLP should be conducted under the Cartesian coordinate for density and view dependent albedo inferences. 
{
\begin{equation}
L_{\text{NLOS}}:(x',y', r, \theta,\phi) \stackrel{R}{\longrightarrow} (x, y, z, \theta, \phi) \stackrel{\text{MLP}}{\longrightarrow} (\sigma, \rho)
\end{equation}
}
Same as NeRF, a key benefit of NeTF is that we no longer need to discretize the scene into a fixed-resolution volume representation. Instead, the deep network representation can provide scene reconstructions at an arbitrary resolution, recovering fine details largely missing in prior art.  


\subsection{Forward Model}

We first reformulate the NLOS reconstruction problem as a forward model under our NeTF representation. Under the confocal setting \cite{2018LCT, 2019FK, 2020MarikoE}, the illumination and detection collocate at the same spot $P(x',y')$ on a relay wall, producing a spherical wave anchored at the spot. The transient $\tau_{\text{iso}}(x', y',t)$ recorded at each spot $P(x', y')$ is the summation of photons that are reflected back at a specific time instant $t$ from the NLOS scene in the 3D half-space $\Omega$ as: 

{
\begin{equation}
\begin{split}
\tau_{\text{iso}}(x',y',t) = \underset{\Omega}{\iiint} \frac{1}{r^4} \rho_{\text{iso}}(x,y,z)g(x', y', x, y, z)\\
\cdot \delta(2\sqrt{(x'-x)^{2}+(y'-y)^{2}+z^{2}}-tc) \di{x}\di{y}\di{z}
\end{split}
\label{isoTau}
\end{equation}
}

\noindent where $c$ is the speed of light, $r$ is the distance between the wall and the NLOS scene as $r = \sqrt{(x'-x)^{2}+(y'-y)^{2}+z^{2}} = tc/2$, and $1 / r^{4}$ is the light fall-off term. $\rho_{\text{iso}}(x,y,z)$ is the albedo of an NLOS point, where $x,y,z$ are the spatial coordinates of the point. Function $g$ models the time-independent effects, including the surface normal, bidirectional reflectance distribution functions (BRDFs), occlusions patterns, etc. The Dirac delta function relates the time of flight $t$ to the distance $r$. 

Notice that function $g$ makes the imaging process nonlinear. To make the problem tractable, previous linear approximation schemes such as \cite{2018LCT} adopt $g=1$ by assuming that the NLOS scene scatters isotropically and that no occlusions occur within the NLOS scene. Such assumptions, however, restrict NLOS scenes to being Lambertian and convex. In contrast, NeTF, by adopting a deep network to model the imaging process, can tackle non-linearity without imposing explicit constraints on $g$. 

We intend to specify how much an NLOS point in the hemisphere contributes to the transient through photons propagation. Consider a detection spot $P(x',y')$ to an NLOS point $Q(r,\theta, \phi)$ in a hemisphere centered at $P$. Recall that the scattering equation \cite{1984Kajiya} serves as the foundation for volume rendering and thereby NeRF. We first derive a photon version of the scattering equation, with more details provided in the supplementary materials. 
 
\begin{figure}[!t]
\centering
\includegraphics[width=0.4\textwidth]{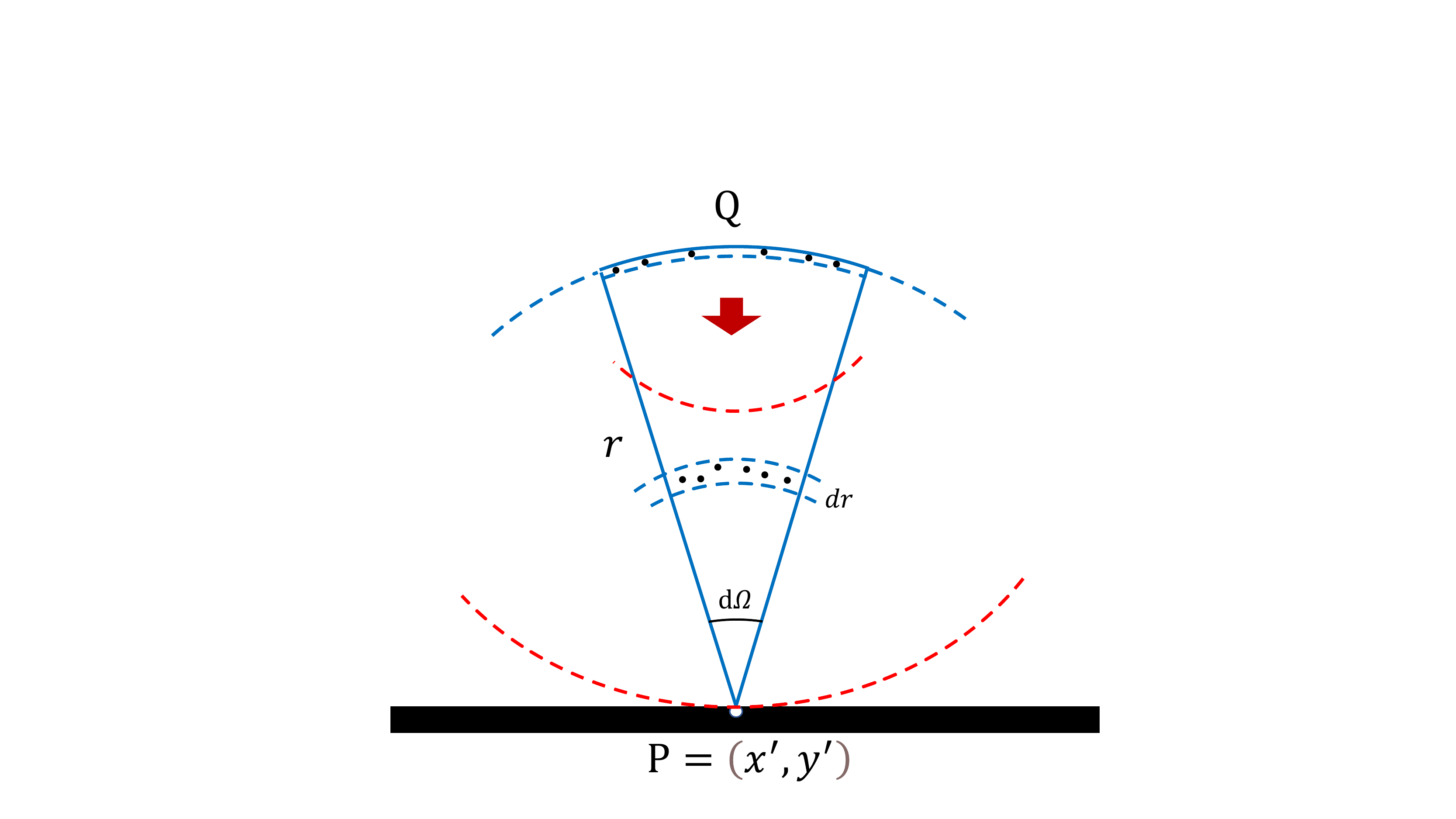}
\caption{The volume rendering model derived under the spherical coordinate system, suitable for processing in NeTF.  
}
\label{forwardScene}
\end{figure}
 

In our NLOS setting, photons travel along spherical wavefronts. When they reach either the relay wall or the hidden surface, they are reflected and then continue to propagate along a hemisphere. We assume that the spot $P$ is a patch with radius $r_0$, and that the location $Q$ in the hidden scene with its neighbors that contribute to $P$ forms a spherical cross-section with radius $r$, thickness $\di{r}$, and a solid angle $\di{\Omega}$. Fig.~\ref{forwardScene} shows that photons travel from $P$ to $Q$ and back from $Q$ to $P$. When $\di{r}$ is sufficiently small, the inner and outer surface areas of the cross-section are $S=r^{2} \di{\Omega}$, and the volume is $S \di{r}$. Recall that $\sigma$ is the density of particles in the cross-section, therefore the number of particles is $\sigma S\di{r}$. Assuming a particle has a radius $a$, the projected area on the surface can then be computed as $A = \pi a^{2}$. We assume that light energy $E$ is attenuated due to particles' absorption and scattering and consequently the energy loss $dE$ can be computed as: 

\begin{equation}
\begin{split}
\di{E} = -\frac{\pi a^{2} \sigma r^{2} \di{\Omega} \di{r}}{r^{2} \di{\Omega}}E = -A \sigma E \di{r}
\end{split}
\label{dE}
\end{equation}

The attenuation coefficient can be computed as $e^{\int_{0}^{r}-A\sigma(r',\theta,\phi) \di{r'}}$ along the radius $r$. Recall that the spot $P$ has a radius $r_0$ and emits radiant energy as a constant $E_P$. Taking integral of Eqn.~\ref{dE}, energy received at $Q$ is defined as: 

\begin{equation}
E_{Q} =  \exp{\left(\int_{0}^{r}-A\sigma(r',\theta,\phi) \di{r'}\right)} E_P \frac{r^2 \cdot \di{\Omega}}{r^2 \cdot 2\pi}
\label{EQ}
\end{equation}

We now consider the reflection at $Q$. Assume that the cross-section is sufficiently thin, e.g., $\di{r}=2a$, the radiant energy at $Q$ attenuated due to absorption and reflection w.r.t. the reflectance $\rho$ can be defined as

\begin{equation}
E_{Q}'(r, \theta, \phi) = A \cdot \sigma(r, \theta, \phi) \cdot 2a \cdot \rho(r, \theta, \phi) \cdot E_{Q}
\label{EQ'}
\end{equation}

On the returning path from $Q$ to $P$, the wavefronts form hemispheres centered at $Q$ with radius $r$. The spot $P$ with area $\pi r_0^2$ receives the photons back to the relay wall, we thus have the energy at $P$ w.r.t. the solid angle $\di{\Omega}$ as: 

\begin{equation}
E_{P}'(r, \theta, \phi) =  \exp{\left(\int_{0}^{r}-A\sigma(r',\theta,\phi) \di{r'}\right)} E_{Q}'\frac{r_0^2 \cdot 2\pi}{ r^2 \cdot 2\pi} 
\label{EP'}
\end{equation}

By taking the integral of Eqn.~\ref{EP'} w.r.t. the solid angle $\di{\Omega}$, we obtain energy received at the detection $P(x', y')$ in the hemisphere at a time instant $t$ as: 

\begin{equation}
   \tau(x', y', t) = \underset{H(x', y';\frac{ct}{2})}{\iint} E_{P}'(r, \theta, \phi) \di{\Omega}
 \label{tauIni}
\end{equation}

Eqn.~\ref{tauIni} serves our forward imaging model. It essentially maps an NLOS point $Q(r, \theta,\phi)$ to a transient $\tau$ detected at a spot $P(x', y')$ on a diffuse surface at a time instant $t$.  For clarity, we abbreviate $\sigma(r, \theta,\phi ; x', y')$ as $\sigma(r, \theta, \phi)$, and $\rho(r, \theta,\phi ; x', y')$ as $\rho(r, \theta, \phi)$. Substituting Eqns.~\ref{EQ}, \ref{EQ'} and \ref{EP'}, Eqn.~\ref{tauIni} is rewritten as:  

\begin{equation}
\begin{split}
    & \tau(x', y', t) = \\ 
    & \Gamma_{0} \underset{H(x', y';\frac{ct}{2})}{\iint} \frac{1}{r^2} \sigma(r, \theta, \phi) \rho(r, \theta, \phi)  exp{\left(2\int_{0}^{r}-A\sigma \di{r'}\right)} \di{\Omega} 
\end{split}
\label{domega}
\end{equation}

\noindent where constant $\Gamma_{0} = {Aar_0^2E_P}/\pi$ is determined by particle radius $a$, initial energy $E_{P}$, and patch radius $r_0$. The integration domain $H(x', y';\frac{ct}{2})$ is a hemisphere centered at $P(x',y')$ on a relay wall, with a radius of $r=ct/2$. $\theta$ and $\phi$ are the elevation and azimuth angles in the viewing direction from $P(x',y')$ to an NLOS point, equivalent to those in the direction of reflection from the NLOS scene. $\rho(r, \theta,\phi ; x', y')$ models view-varying BRDFs of the NLOS scene. $e^{2\int_{0}^{r}-A\sigma(r',\theta,\phi) \di{r'}}$ is an exponential actuation coefficient and reveals the visibility of an NLOS point with respect to varying detection spots $P(x',y')$. Since $d\Omega = \sin{\theta}\di \theta \di \phi$, we have: 
    
\begin{equation}
\begin{split}
    \tau(x', y', t) = \Gamma_{0} \underset{H(x',y';\frac{ct}{2})}{\iint} \frac{\sin{\theta}}{r^2} \sigma(r, \theta, \phi) \rho(r, \theta, \phi) \cdot \\
    exp{\left(2\int_{0}^{r}-A\sigma \di{r'}\right)} \di{\theta} \di{\phi}
\end{split}
\label{tau}
\end{equation}


Recall that the forward plenotpic transient field model in Eqn.~\ref{tau} is computationally expensive if we use the MLP for training. If we further assume that the NLOS scene is all opaque and does not exhibit self-occlusions, we can further simplify the formulation to:

{
\begin{equation}
\begin{split}
    \tau(x', y',t) = \Gamma_{0} \underset{H(x', y';\frac{ct}{2})}{\iint} \frac{\sin{\theta}}{r^{2}} \sigma(r, \theta,\phi) \rho(r, \theta,\phi)  \di{\theta} \di{\phi}
\end{split}
\label{tauSim}
\end{equation}
}

Such a formulation reduces computations and is used in examples of Figs.~\ref{TwoStage},~\ref{LucyStatue},~\ref{SimulatedResult},~\ref{DLCT}, and~\ref{RealResult} to accelerate processing. Its downside though is that, unlike Eqn.~\ref{tau} , it cannot handle occlusions. For scenes that contain heavy occlusion, we use Eqn.~\ref{tau} , e.g., in Fig.~\ref{Semioccluded}. 

Finally, it is noting that although both NeRF and our NeTF derive the forward model based on volume rendering, NeRF models how a ray propagates along a line (i.e., with a cylinder between two points) whereas NeTF models spherical wavefront propagation (i.e., with a cone model that accounts for attenuation). In addition, the volume rendering model used in NeRF only considers one-way accumulation, i.e., how light travels through light emitting particles towards the camera sensor. In contrast, our NeTF adopts a two-way propagation model, i.e., how light illuminates the scene and how scenes illuminate the wall.  

 
\subsection{Differentiable Rendering}
\label{Differential Rendering}

Our forward model is differentiable, we can therefore numerically compute the continuous integral Eqn.~\ref{tauSim} using quadrature, as:

\begin{equation}
\begin{aligned}
\tau(x',y',t) &= \frac{\Delta \theta \Delta \phi}{r^2} \sum_{i,j} \sin{(\theta_{ij})} \sigma(r,\theta_{ij},\phi_{ij}) \rho( r, \theta_{ij},\phi_{ij}) 
\end{aligned}
\end{equation}

 
\noindent where $Q(r,\theta_{ij},\phi_{ij})$ are scene points uniformly sampled along the hemispherical rays. We transform these points into their corresponding Cartesian coordinates as inputs to the MLP. The network outputs the density and reflectance at each point. We then sum all the outputs as neural transient fields from the transients. 

We optimize our NeTF by minimizing the following $l_2$-norm loss function as the difference between the predicted $\tau(x',y',t)$ and measured $\tau_m(x',y',t)$ transients:

\begin{equation}
L =\sum_{x',y',t} (\tau_m(x',y',t) - \tau(x',y',t))^2
\end{equation}

Notice that the use of MLP allows minimizing arbitrary losses as long as they are differentiable with respect to our prediction $\tau(x',y',t)$, although $l_2$-norm is most commonly adopted same as in NeRF. 








\begin{figure}[!t]
\centering
\includegraphics[width=0.48\textwidth]{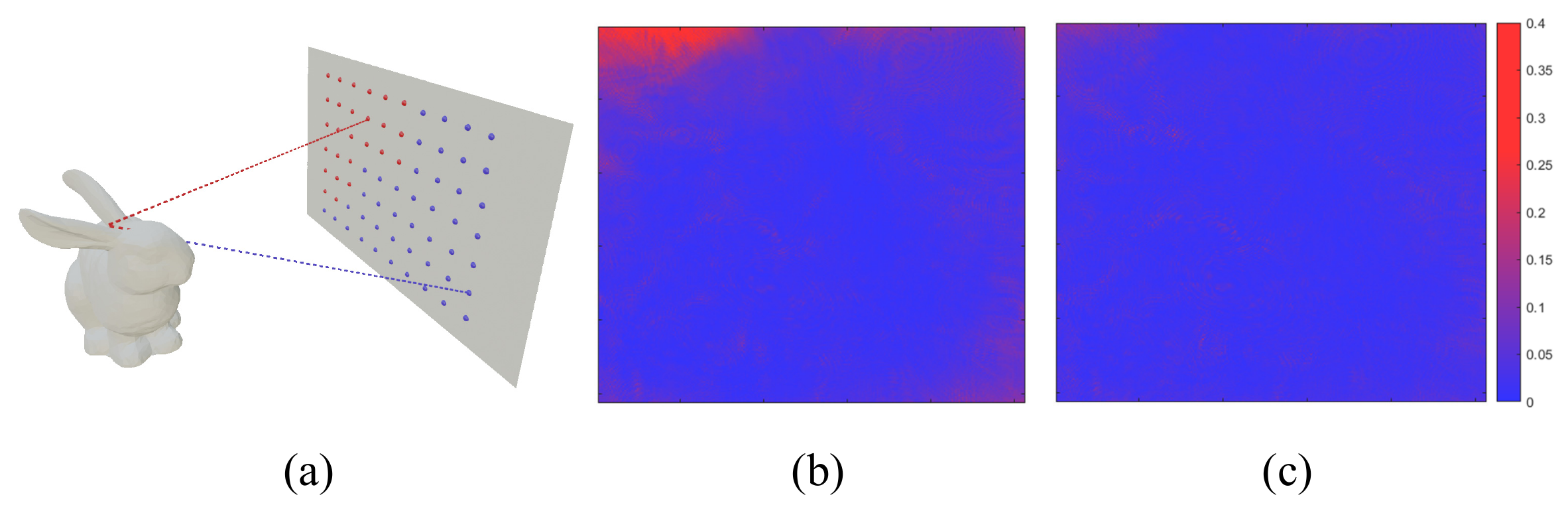}
\caption{We use the loss induced from the first stage in the training process to guide resampling. At each spot on the relay wall (a), we measure the loss (b) and apply our second stage for resampling. (c) shows the final loss after applying resampling.}
\label{Loss}
\end{figure}

\section{Neural Transient Fields Optimization}

Our NeTF forward model allows modeling the plenoptic transient field using an MLP. However, data acquired by NeTF are quite different from those in NeRF. In NeRF, a dense set of high resolution images is generally required to produce satisfactory density estimation and view interpolation. Under the dense viewpoint setting, the problem of occlusions is less significant as there will be a sufficient number of views capturing the occluded point to ensure reliable reconstruction. In NeTF, however, our SPAD only captures a sparse set of spots on the wall and an occluded point may be captured only from a very small number of viewpoints (spots). Consequently, occlusion can lead to strong reconstruction artifacts if not handled properly. We develop a two-stage training strategy along with a hierarchical sampling technique to address this sampling bias. 

\subsection{Two-stage Training}
\begin{figure}[!t]
\centering
\includegraphics[width=0.48\textwidth]{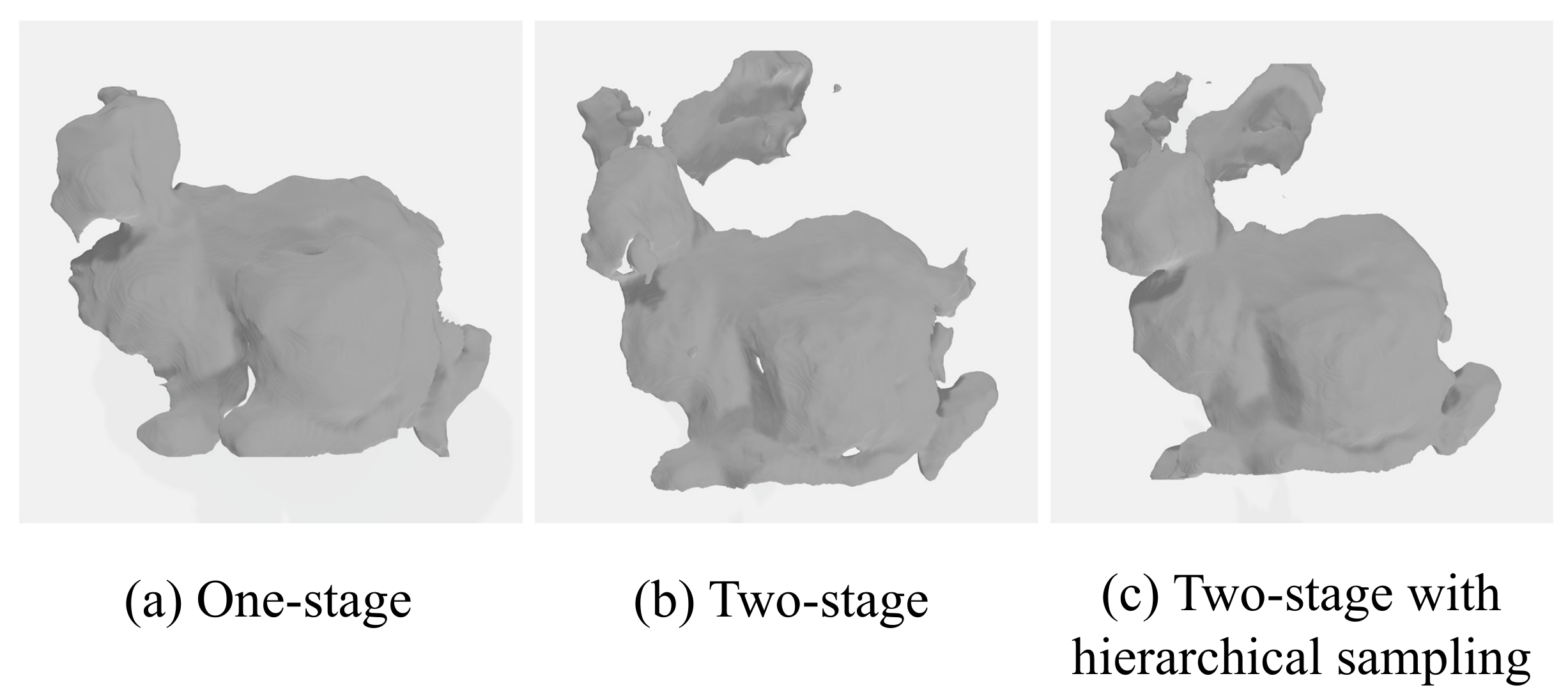}
\caption{From left to right: recovered results using our one-stage training, two-stage training without hierarchical sampling, and two-stage with hierarchical sampling. One-stage training, same as most prior art (Fig.~\ref{SimulatedResult}), fails to recover the second ear of the bunny. Our two-stage schemes manage to recover both ears of the bunny. With hierarchical sampling, NeTF further improves reconstruction with more complete shape and more accurate silhouettes.}
\label{TwoStage}
\end{figure}

We observe that the sampling bias resembles the long-tailed classification problem in machine learning. A conventional solution is to resample the dataset to achieve a more balanced distribution by over-sampling the minority classes \cite{2020Kang}. We therefore adopt a two-stage training strategy. We first conduct training using all samples to obtain an initial reconstruction. We then calculate the loss function between the predicted and measured transients at every measuring spot on the relay wall. We observe spots that correspond to a high loss imply undersampling and should incur more samples. We therefore normalize the calculated loss to form a probability density function (PDF). Next, we resample the detection spots using the PDF: a higher loss corresponds a higher PDF and should be more densely sampled. We thus use this sampling scheme to build a new training dataset and then retrain our network to refine reconstruction. The bunny scene (Fig.~\ref{Loss}) shows a sample loss map (and therefore resampling density map) with 256x256 sampling spots on the relay wall, with red implying a higher loss (and thus high PDF in subsequent sampling) and blue a low one. Using two stage training, we manage to significantly reduce the loss near the upper left spots on the wall through which the ears of the bunny should be observed. Consequently, our two-stage training manages to recover bunny ears largely missing in one-stage training and in prior art (Figs.~\ref{TwoStage} and ~\ref{SimulatedResult}).




The two-stage training process provides a viable solution to tackle imbalanced sampling for achieving more accurate reconstruction. Fig.~\ref{TwoStage} demonstrates the corresponding reconstruction results with and without using the second stage. Results with the second stage manage to recover many fine details largely missing from the first stage, e.g., the ear of Bunny and details of the abdomen regions.

\subsection{Hierarchical Sampling}

Denser samples produce higher quality reconstruction. At the same time, they lead to a much higher computation overhead. For example, by uniformly sampling $L$ hemisperical wavefronts at each detection spot and $N^2$ scene points on each $L$, the resulting training process requires a computational complexity of $O(N^2L)$. We observe that under the confocal setting, spherical wavefronts only intersect with a very small portion of the NLOS scene. These wavefronts tend to converge at specific patches and contribute greatly to the final integral where the contributions from the rest are negligible.

Note that the hierarchical sampling scheme in NeTF is different from NeRF: NeRF calculates the integral along a ray, i.e., using 1D sampling, whereas NeTF on a hemisphere, i.e., using 2D sampling. To make this problem tractable, we develop a \textit{coarse}-to-\textit{fine} sampling scheme. Specifically, we first sample $N^2_c$ uniform scene points in the hemisphere and evaluate our \textit{coarse} network with the estimated PDF $k(\theta,\phi)$. We then employ Metropolis-Hastings algorithm and conditional Gaussian distribution for state transition of Markov chain to produce a \textit{fine} sampling of $N_f$ scene points as $K(\theta^{f}_{ij},\phi^{f}_{ij})$ along the hemispherical wavefronts that intersect with the NLOS scene. We finally combine the coarse and fine samples $N_c^2 + N_f$ to re-evaluate the reconstruction quality from our \textit{fine} network. 

Specifically, 
\begin{equation}
\tau(x',y',t) = \frac{\tau_c(x',y',t) + \tau_f(x',y',t)}{2}
\end{equation}

\noindent where $\tau_c(x',y',t)$ is the integral with the coarse samples $N_c^2$, and $\tau_f(x',y',t)$ is estimated with samples $N_f$ from MCMC, as:

\begin{equation}
\begin{aligned}
&\tau_f(x',y',t) = \frac{1}{r^4} \sum_{i,j} \frac{ \sigma(r, \theta^{f}_{ij},\phi^{f}_{ij}) \rho(r, \theta^{f}_{ij},\phi^{f}_{ij})}{K(\theta^{f}_{ij},\phi^{f}_{ij})}
\end{aligned}
\end{equation}

It is important to note that our hierarchical sampling is intrinsically differentiable (\ref{Differential Rendering}). Previous volume-based methods, e.g., \cite{2012Velten, 2018LCT, 2020DLCT, 2019FK}, in theory can also apply such a hierarchical sampling technique to refine their reconstruction. In reality, these methods use an explicit volumetric representation with a fixed resolution, making resampling on the hemisphere intractable.

\begin{figure}[!t]
\centering
\includegraphics[width=0.47\textwidth]{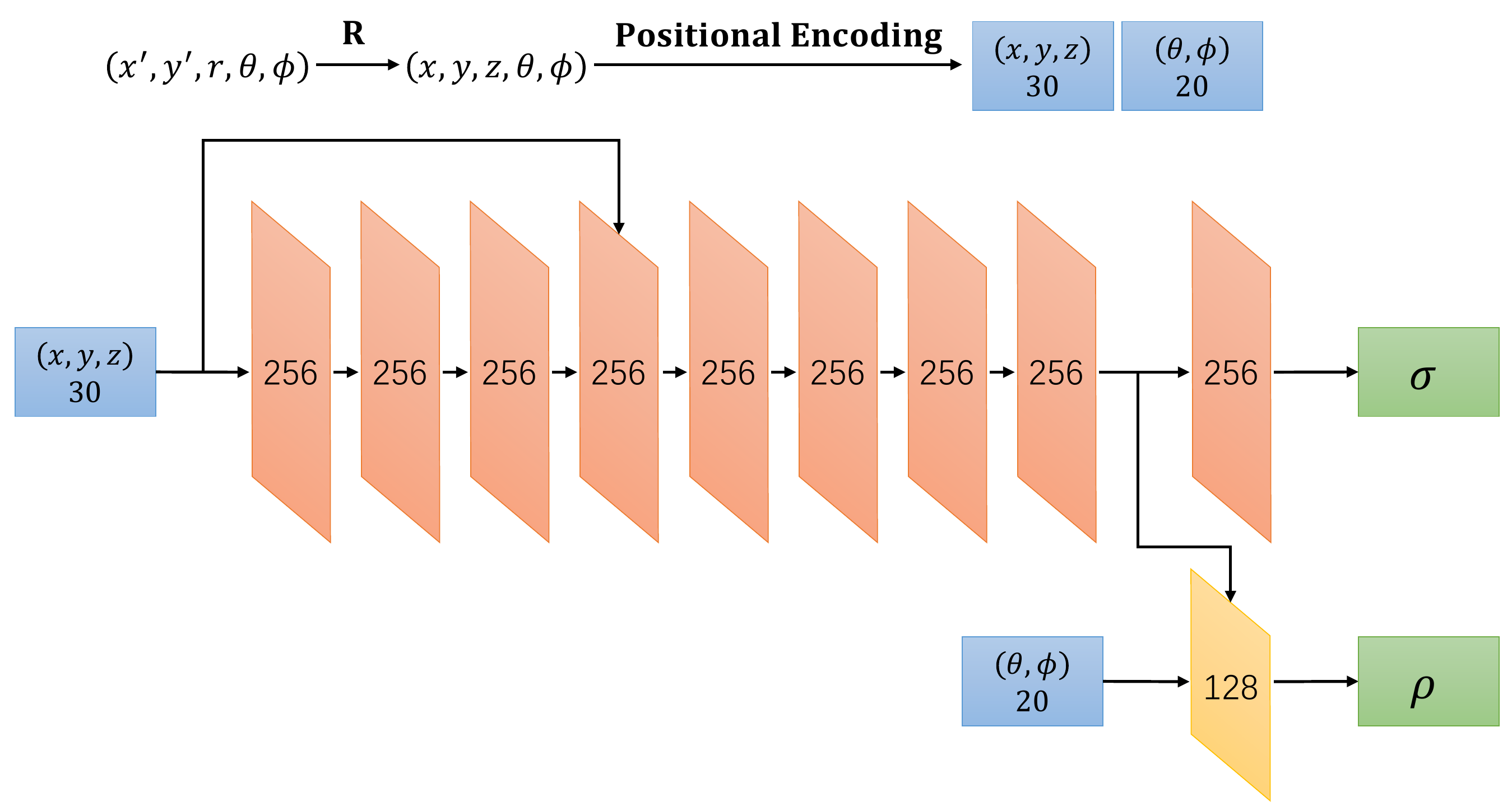}
\caption{NeTF network architecture: we adopt an MLP structure analogous to the one used in NeRF. The key differences are (1) NeTF uses ReLU vs. NeRF uses sigmoid and (2) the last four layers in NeRF are simplified to one layer.}
\label{MLP}
\end{figure}

\begin{figure*}[htbp]
\centering
\includegraphics[width=0.81\textwidth]{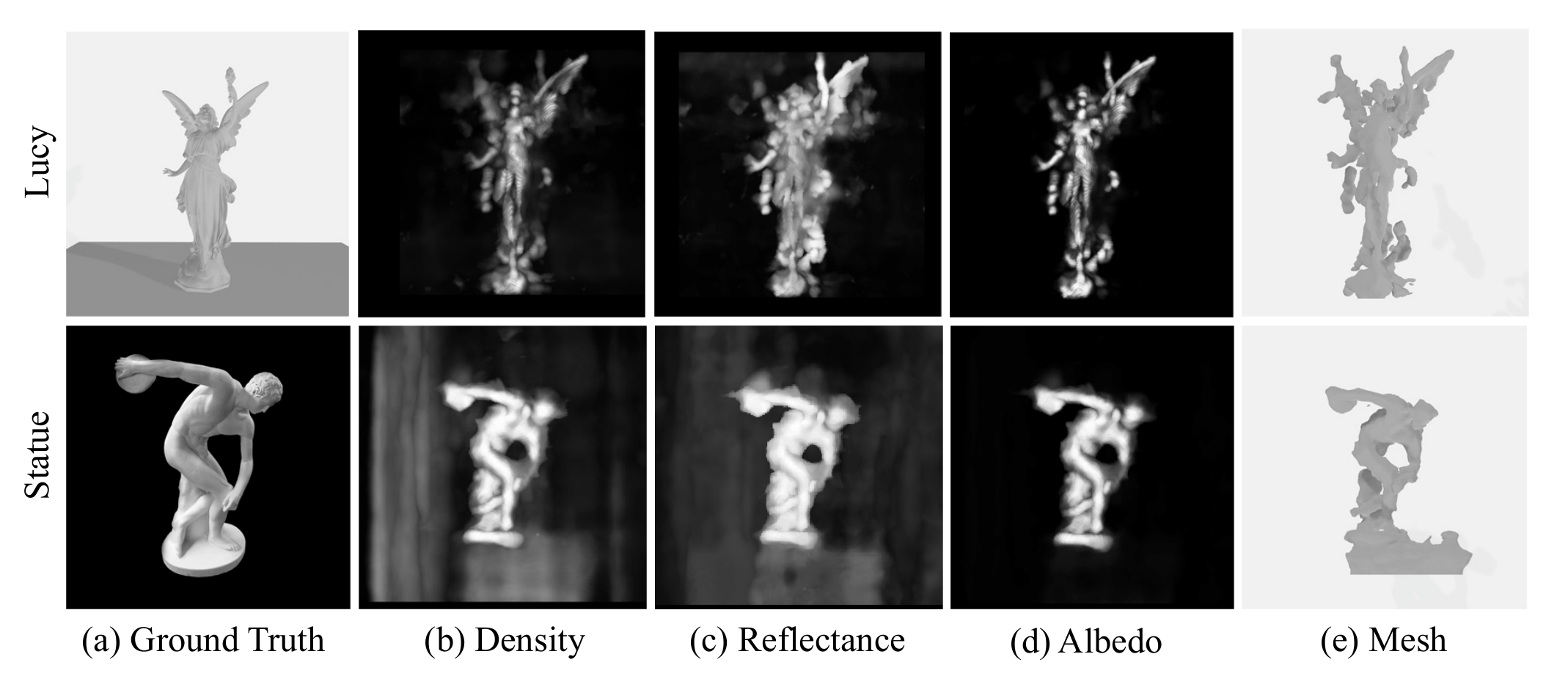}
\caption{From left to right: the ground truth, the recovered volume density, reflectance, albedo, and 3D mesh reconstruction using NeTF. Top shows the results on the Lucy model and bottom on the Statue. }
\label{LucyStatue}
\end{figure*}

\section{Experimental Results}
In this section, we discuss our NeTF implementation and experimental validations.

\begin{figure*}[htbp]
\centering
\includegraphics[width=0.81\textwidth]{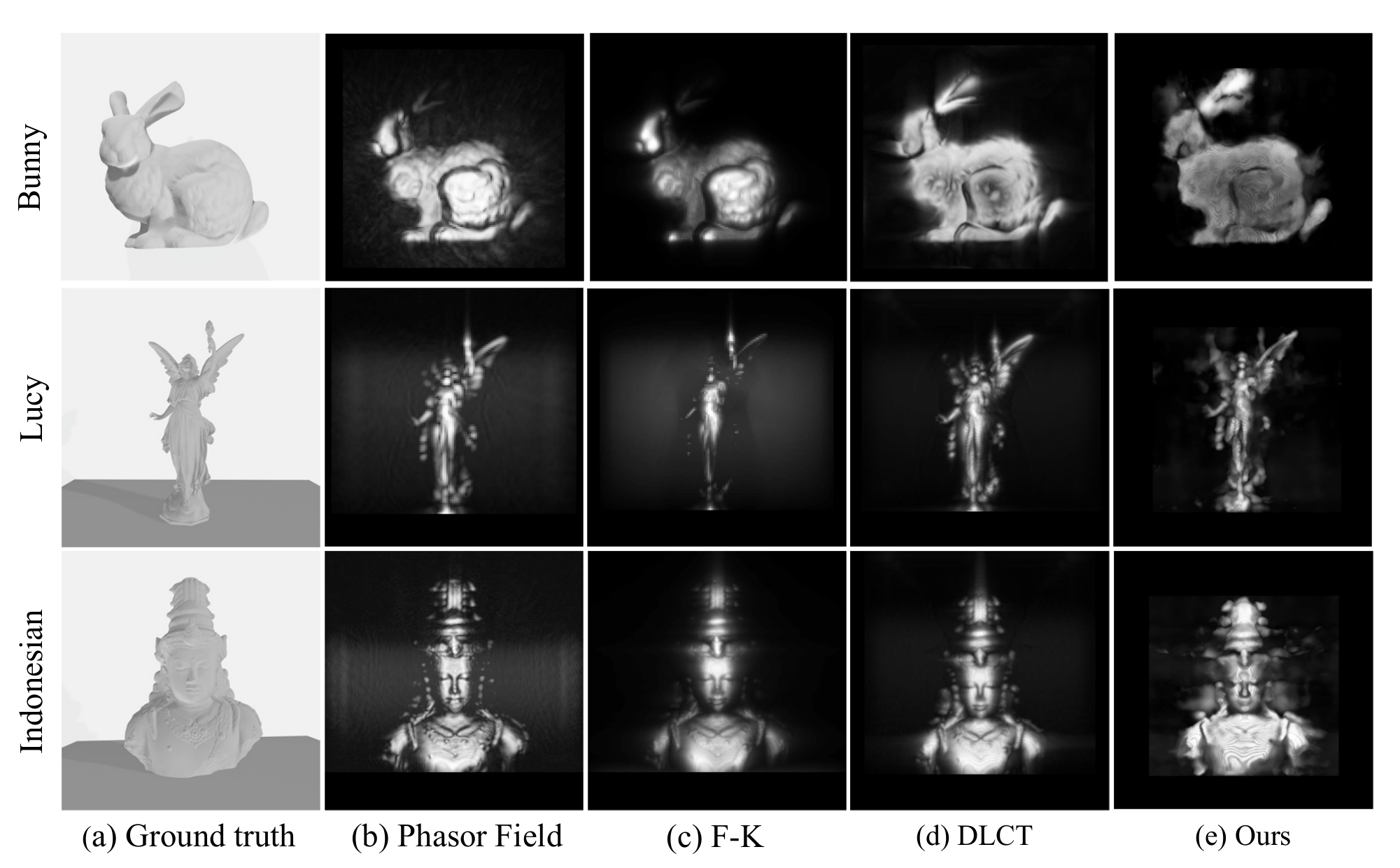}
\caption{Comparisons on simulated NLOS data. NeTF achieves comparable reconstructions as SOTA and further manages to recover challenging geometry such as the ear of Bunny, the wing of Lucy and the crown of Indonesian.}
\label{SimulatedResult}
\end{figure*}
 
\begin{figure*}[htbp]
\centering
\includegraphics[width=0.75\textwidth]{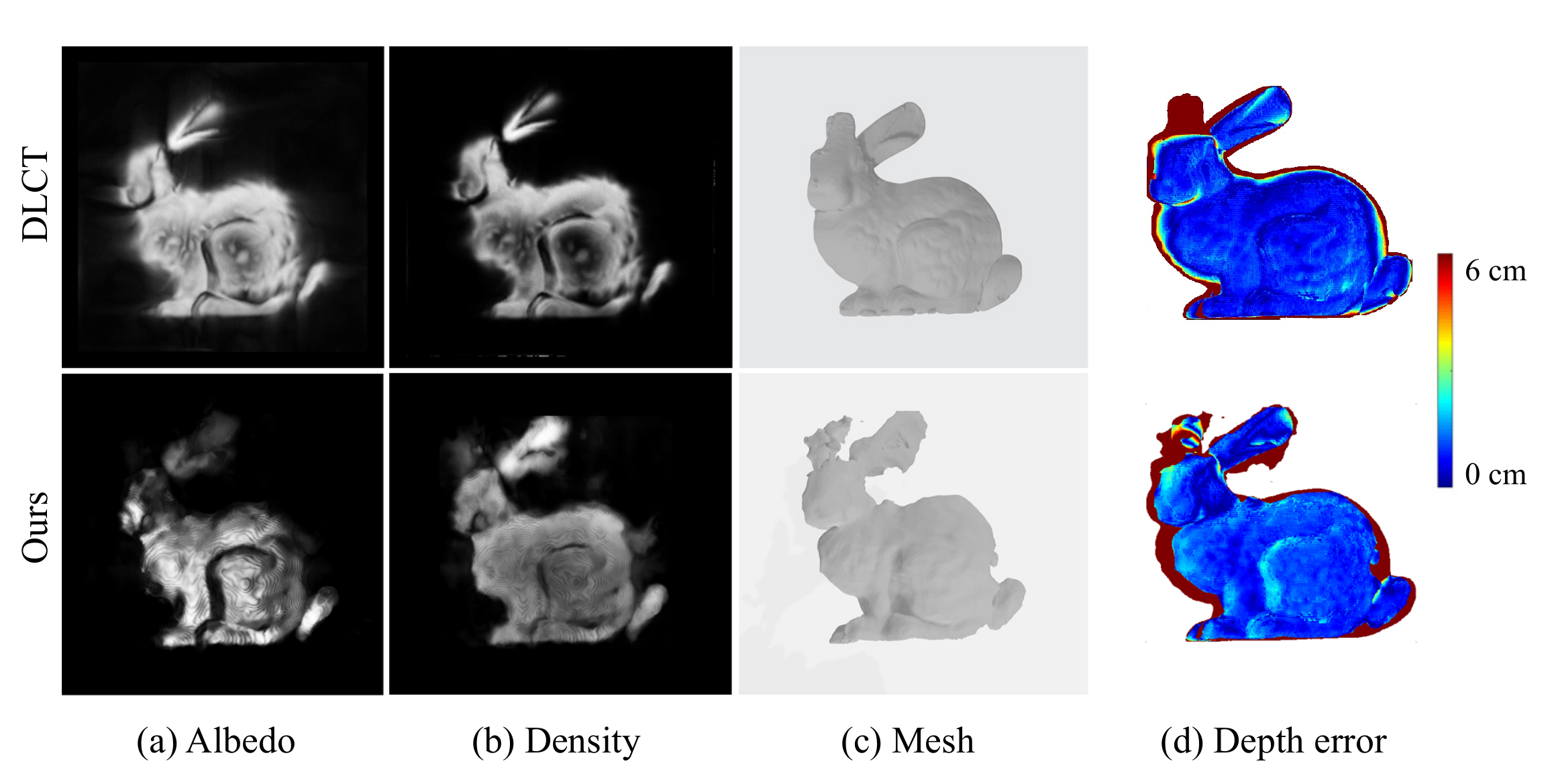}
\caption{Comparisons between NeTF and DLCT on the Bunny scene. Both methods manage to acquire the overall geometry yet DLCT misses one ear whereas NeTF captures both. In its own implementation \cite{2020DLCT}, DLCT further uses the mask (silhouettes) of the bunny to further improve reconstruction to obtain the final mesh.  }
\label{DLCT}
\end{figure*}

\begin{figure*}[htbp]
\centering
\includegraphics[width=0.79\textwidth]{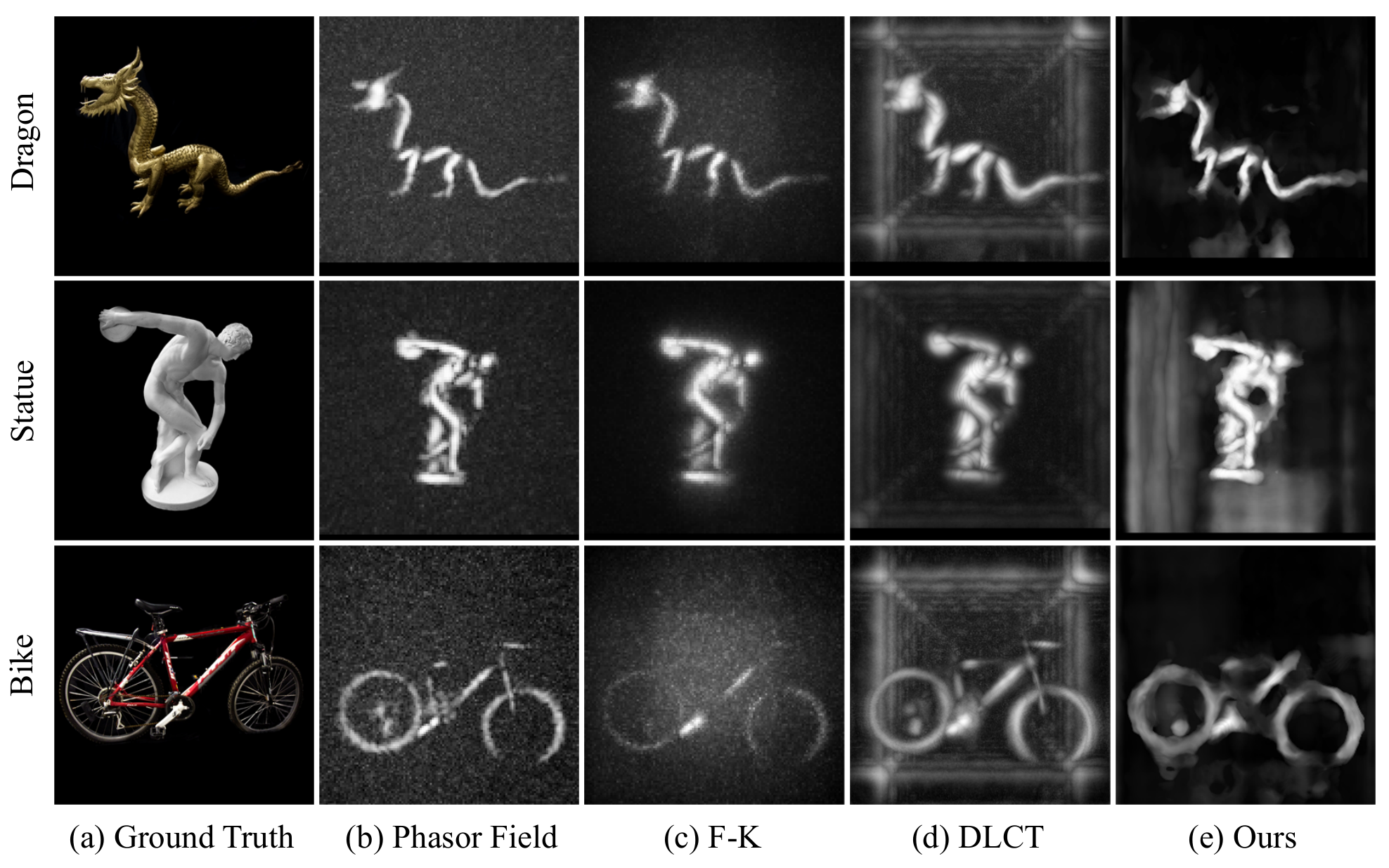}
\caption{Comparisons on real NLOS data. (a) The ground truth shows a photography of the hidden object. (b-c) show results using various techniques. The Bike dataset exhibit heterogeneous materials and complex topology and are particularly challenging. NeTF produces comparable reconstructions to SOTA. In particular, same as other neural modeling methods such as NeRF, NeTF reconstruction incurs much lower noise than SOTA. } 
\label{RealResult}
\end{figure*}

\subsection{MLP Settings}
We train the NeTF using an MLP. Fig.~\ref{MLP} shows the structure of our MLP. Analogous to NeRF \cite{2020NERF}, we construct a fully connected network with nine 256-channel layers, and with one 128-channel layer. We use ReLU activations for all the layers. We transform the spherical coordinates of sampling points into their Cartesian coordinates $(x,y,z,\theta,\phi)$, and feed them into the MLP. We predict the volume density as a function of only position, and the view-dependent reflectance as a function of both position and direction. 

We first normalize the spatial coordinates $(x,y,z)$ and the viewing direction $(\theta,\phi)$ to range [-1, 1]. Next, we apply the positional encoding (PE) technique and map each input from 1 dimension onto a 10-dimensional Fourier domain to represent high-frequency variation in geometry and reflectance. Our MLP then processes the coordinates $(x,y,z)$ as inputs with eight 256-channel layers and outputs a 256-dimensional feature vector. Note that we also concatenate $(x,y,z)$ with the fourth layer for skip connection. This feature vector is passed to an additional 256-channel layer and produces $\sigma$. Simultaneously, the feature vector is concatenated with the direction $(\theta,\phi)$ and passed to the 128-channel layer for reflectance $\rho$.
 
Under the NLOS setting, we consider a batch size of 1 to 4 transients and employ $32 \times 32$ or $64 \times 64$ samples for both uniform sampling $N_c^2$ and MCMC sampling $N_f$ on the hemisphere. We adopt the Adam optimizer \cite{2015Adam} with hyperparameters $\beta_{1}=0.9$, and $\epsilon = \num{1e-7}$. In our experiments, we use a learning rate that begins at $\num{1e-3}$ and decays exponentially to $\num{1e-4}$ through the optimization. The training time of NeTF shares certain similarities to NeRF. In NeRF, the training cost depends on how densely we sample along each ray. In NeTF, it depends on two factors: how densely we sample the radius of the hemisphere (i.e., the number of layers) and how densely we sample each layer/hemisphere. For the Bunny scene, on a single GeForce RTX 3090 GPU the training takes 10 hours using 200 layers with $32\times32$ samples on each layer (5 epochs, batchsize 4). The training time quadruples with the same number of layers but at $64\times64$ samples.

\begin{figure*}[htbp]
\centering
\includegraphics[width=0.66\textwidth]{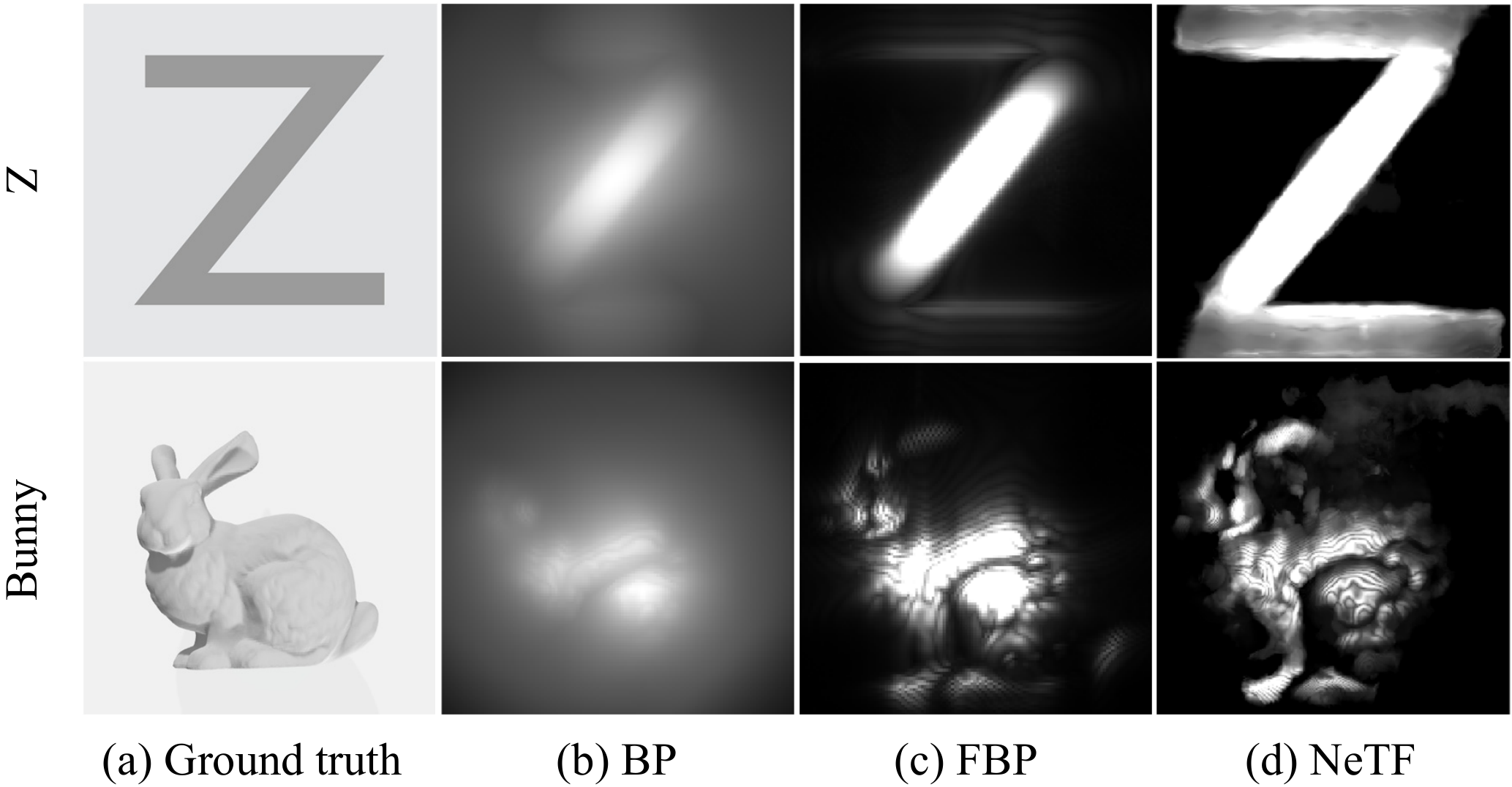}
\caption{Visual Comparisons of NLOS reconstructions by NeTF and SOTA under the non-confocal setting. NeTF manages to recover clearer silhouettes than SOTA.}
\label{nonconfocal}
\end{figure*}

\begin{figure}[!t]
\centering
\includegraphics[width=1\columnwidth]{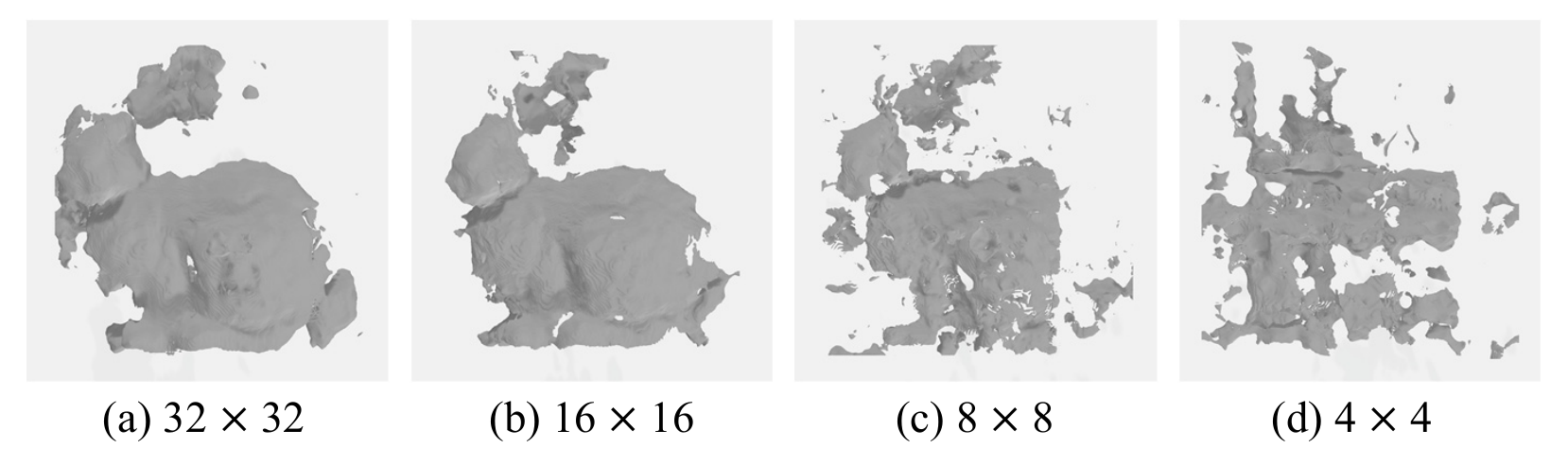}
\caption{Low Data Resolution: Mesh reconstructed from simulated transients of Bunny at $32\times32$, $16\times16$, $8\times8$, and $4 \times4$ spots on the wall. Even at a very low resolution of $8 \times8$, our NeTF produces reasonable reconstructions. }
\label{LowDataResolution}
\end{figure}
 
\subsection{Validations}
We have validated our approach on two public NLOS datasets: a simulated ZNLOS dataset \cite{2014Jarabo, 2019Galindo}, and a real Stanford dataset \cite{2019FK}. ZNLOS consists of multi-bounce transients of synthetic objects that are 0.5 m from the relay wall. The transients have a temporal resolution of 512 time bins with a width of 10 ps and a spatial resolution of 256 $\times$ 256 pixels. The Stanford dataset captures transients measured in real scenes that are 1.0 m away from the relay wall. The transients in this dataset have a temporal resolution of 512 time bins with a width of 32 ps and a spatial resolution of 512 × 512 or 64 × 64 pixels. We conduct quantitative and qualitative comparisons between NeTF and the state-of-the-art (SOTA) methods. 

\textbf{Qualitative Comparisons.}
On ZNLOS, we experiment on several simulated hidden objects: Bunny, Lucy, and Indonesian at a spatial resolution of $256 \times 256$ pixels that correspond to an area of size 1 m $\times$ 1 m on the relay wall. All three models are diffuse; Bunny does not contain a floor, but Lucy and Indonesian do. 
On the Stanford dataset, we have experimented on three real hidden objects with different materials: a diffuse Statue, a glossy Dragon, and a metal Bike. Their spatial resolution is of $512 \times 512$ spots and we down-sample them to $256 \times 256$, same as in \cite{2020DLCT} ($128\times128$ in \cite{2019FK, 2019Liu}).

Fig.~\ref{LucyStatue} illustrates our results. Our NeTF outputs a volume density map $\sigma$ and a directional reflectance map $\rho$ of ZNLOS Bunny and Stanford Statue. From these two maps we can produce volumetric albedo, and reconstruct a 3D mesh of hidden objects. By sampling $256\times256$ transients, our NeTF produces high quality reconstructions of objects with complex textures (e.g., Lucy). The density and reflectance maps of both Statue and Lucy, and the volumetric albedo produce much less error. We can then apply the Marching Cubes algorithm to further convert the volume to surfaces.

Fig.~\ref{SimulatedResult} shows the comparisons with three most broadly adopted volume-based methods \cite{2019FK, 2019Liu, 2020DLCT}. We compare the projected volumes to 2D maps of Indonesian, Lucy and Bunny. Note that the results from \cite{2019FK, 2019Liu} correspond to volumetric albedos whereas ours includes both the density and the albedo maps. We also show the normal volume using \cite{2020DLCT}.  The recovered volume maps on these hidden objects demonstrate NeTF achieves comparable reconstruction quality to SOTA. NeTF, however, can tackle challenging geometry, e.g., the ear of Bunny, the wing of Lucy and the head of Indonesian that are partially missing using prior art. The Phasor Field technique achieves the best performance on Indonesian but still misses the ear on bunny and wing on Lucy. This implies that such geometry may cast additional challenges to wave-based techniques but can be potentially recovered via volume reconstruction. 
 
DLCT \cite{2020DLCT} produces comparable results to NeTF on Bunny and Indonesian. On Bunny, both methods manage to acquire the overall geometry yet DLCT misses one ear whereas NeTF captures both. In Fig.~\ref{SimulatedResult}, and~\ref{DLCT}, DLCT further uses the mask (silhouettes) of the bunny to obtain the final mesh. The use of the mask can recover the shape (depth) of both ears but the geometry of the second ear is still incorrect. NeTF, in contrast, manages to recover both ears of Bunny. Similar reconstructions can be observed on Lucy. Fig.~\ref{DLCT} shows in-depth comparisons between NeTF and DLCT for Bunny on the recovered albedo, normal (density in our results), and mesh reconstruction. Our method preserves fine details but is slightly more noisy, as shown in the depth error. A similar phenomenon is observed on NeRF for multi-view 3D reconstruction where the noise can be potentially filtered.  
 
On the real Stanford dataset, Fig.~\ref{RealResult} compares NeTF vs. SOTA for the glossy Dragon, diffuse Statue, and metal Bike. On Dragon and Statue where view-dependency is relatively small, NeTF and SOTA produce comparable results, although NeTF manages to better preserve high frequency features such as occluding edges. On the challenging Bike scene, NeTF achieves a similar performance to \cite{2019Liu}. For DLCT, the reconstructed mesh exhibits adhesion between different parts, whereas reconstruction produced by NeTF manages to separate these parts.


We have further tested robustness and efficiency of NeTF under non-confocal settings as well as on low resolution transient inputs. Fig.~\ref{nonconfocal} shows the density, albedo and mesh reconstruction. NeTF produces reasonable estimations to the ground truth and significantly higher quality reconstructions compared with the results from BP 
and FBP\cite{2012Velten}. Fig.~\ref{LowDataResolution} shows the NeTF results with down-sampled measurements from $256\times256$ spots to $32\times32$, $16\times16$, $8\times8$ and even $4\times4$. Even with very sparse sampling spots ($16\times16$ and $8\times8$), NeTF produces reasonable reconstructions without any prior. To test our approach under the non-confocal setting, we test on two additional objects from ZNLOS, i.e., the letter Z and the bunny, and their transients simulated under non-confocal setups.

\begin{figure}[!t]
\centering
\includegraphics[width=0.95\columnwidth]{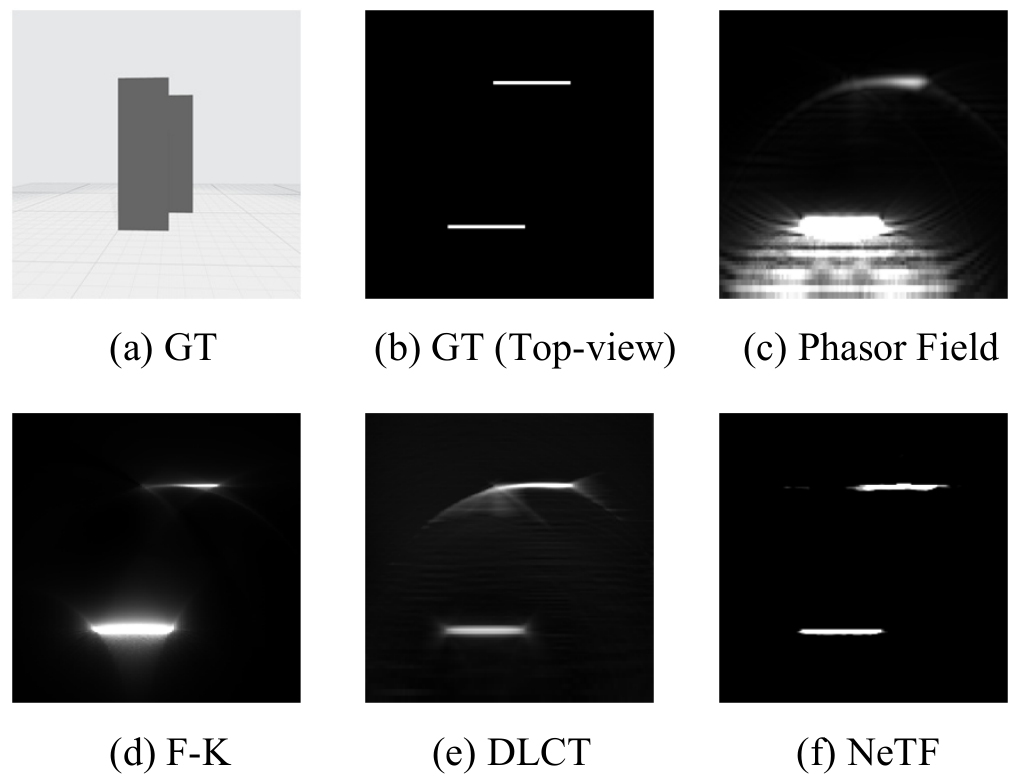}
\caption{Visual comparisons of NLOS reconstructions on the semi-occluded scene. (a) and (b) show the frontal and top-down views. Closest to NeTF is DLCT, which manages to recover the front plane but produces high errors on the back plane.}
\label{Semioccluded}
\end{figure}

To further test the robustness of NeTF vs. SOTA on occlusions, we experiment on a semi-occluded scene from ZNLOS using Eqn.~\ref{tau}. Fig.~\ref{Semioccluded} shows the frontal and top-viewed albedo maps of the reconstruction. Phasor Field is most sensitive to occlusions whereas DLCT and F-K can only recover one plane at a high accuracy. NeTF produces sharper edges of both front and back planes.

\textbf{Quantitative Comparisons.}
Table~\ref{table1} and~\ref{table2} show that NeTF achieves comparable accuracy as the state-of-the-art (SOTA) in terms of Mean Absolute Error (MAE), demonstrating the feasibility and efficacy of deep neural networks in NLOS under both confocal and non-confocal settings. Under the MAE metric, the gain using NeTF does not seem significant. However, MAE does not fully reflect the reconstruction quality: our experiments have further revealed that NeTF can more robustly handle silhouettes and semi-occlusions, as shown in Fig.~\ref{nonconfocal} and ~\ref{Semioccluded}.

\section{Conclusion and Future Work}
We have presented a novel neural modeling framework called the Neural Transient Field (NeTF) for NLOS imaging. Similar to the recent Neural Radiance Field that seeks to use a multi-layer perception (MLP) to represent the 5D radiance function, NeTF recovers the 5D transient function in both spatial location and direction. Different from NeRF, the input training data are parametrized on the spherical wavefronts in NeTF rather than along lines (rays) as in NeRF. We have hence formulated the NLOS process under spherical coordinates, analogous to volume rendering under Cartesian coordinates. Another unique characteristic of our NeTF solution is the use of Markov chain Monte Carlo (MCMC) to account for sparse and unbalanced sampling in NeTF. MCMC enables more reliable volume density estimation and produces more accurate shape estimation by recovering details caused by occlusions and non-uniform albedo. Our experiments on both synthetic and real data demonstrate the benefits of NeTF over SOTA in both robustness and accuracy. 

\begin{table}[!t]
\renewcommand\arraystretch{1.3}
\centering
\caption{Reconstruction error using NeTF vs. SOTA on three confocal NLOS datasets measured by MAE. NeTF achieves comparable performance in MAE. Notice though MAE does not fully reflect the reconstruction quality: for example, Phasor Field produces the highest MAE on Indonesian, indicating lowest reconstruction quality; yet it manages to recover many fine details largely missing in F-K and DLCT, as shown in Fig.~\ref{SimulatedResult}.}
\begin{tabular}{|c|c|c|c|} 
\hline
MAE & Bunny & Lucy & Indonesian\\ 
\hline      
Phasor Field & 2.89 cm & 1.36 cm & 1.69 cm \\
\hline
F-K  & 2.43 cm & 2.05 cm & 0.61 cm\\
\hline
DLCT & 2.38 cm & 0.23 cm & 0.30 cm\\
\hline
NeTF & 2.65 cm & 1.05 cm & 0.31 cm\\
\hline
\end{tabular}
\label{table1}

\end{table}

\begin{table}[!t]
\renewcommand\arraystretch{1.3}
\centering
\caption{Reconstruction error using NeTF vs. SOTA on two non-confocal NLOS datasets measured by MAE. Same as in Table~\ref{table1}, we observe that low MAE does not sufficiently reflect reconstruction quality, e.g., on the Z letter scene, NeTF performs slightly worse than FBP in MAE but better preserves the silhouettes, as shown in Fig.~\ref{nonconfocal}.}
    \begin{tabular}{|c|c|c|} 
    \hline
    MAE & Bunny (non-confocal) & Z (non-confocal) \\ 
    \hline
    BP & 7.02 cm & 3.21 cm \\
    \hline
    FBP  & 3.77 cm & 0.46 cm \\
    \hline
    NeTF & 7.45 cm & 0.60 cm \\
    \hline
    \end{tabular}
    \label{table2}

\end{table}

Same as NeRF, the final reconstruction of the hidden scene corresponds to a 3D density volume, implicitly represented by the MLP. Recovering the actual shape requires mapping the volume to surfaces, e.g., by thresholding followed by Marching Cubes. Such brute-force implementations may lead to noise on smooth surfaces. There are a number of emerging neural modeling techniques that can potentially provide smooth reconstructions, by imposing shape priors \cite{yariv2020multiview}. In general, learning-based techniques (including NeRF and NeTF), in their current forms, are still substantially more computationally expensive than previous optimization techniques, although we observe a large number of emerging acceleration schemes. More importantly, NeTF demonstrates that deep learning provides an alternative and potentially feasible solution to a broader class of inverse imaging problems. There are also several acceleration schemes, e.g., using results from SOTA to initialize the network and then conduct training. It is our immediate future work to investigate how to integrate such approaches into our NeTF framework. Our current approach does not separately treat the confocal and non-confocal setups. Analogous to multi-view stereo vs. photometric stereo, it may be possible to tailor solutions such as ~\cite{bi2020neural} on top of NeTF to separately handle different settings.

\ifpeerreview \else
\section*{Acknowledgments}
The authors thank anonymous reviewers for their valuable feedback. We also thank Minye Wu and Huangjie Yu for their helpful discussions. This work was supported by NSFC (grant nos. 61976138 and 61977047), and STCSM (2015F0203-000-06).
\fi

\bibliographystyle{IEEEtran}
\normalem


\nocite{*}


\ifpeerreview \else



\vspace{-10mm}
\begin{IEEEbiography}
[{\includegraphics[width=1in,height=1.25in,clip,keepaspectratio]{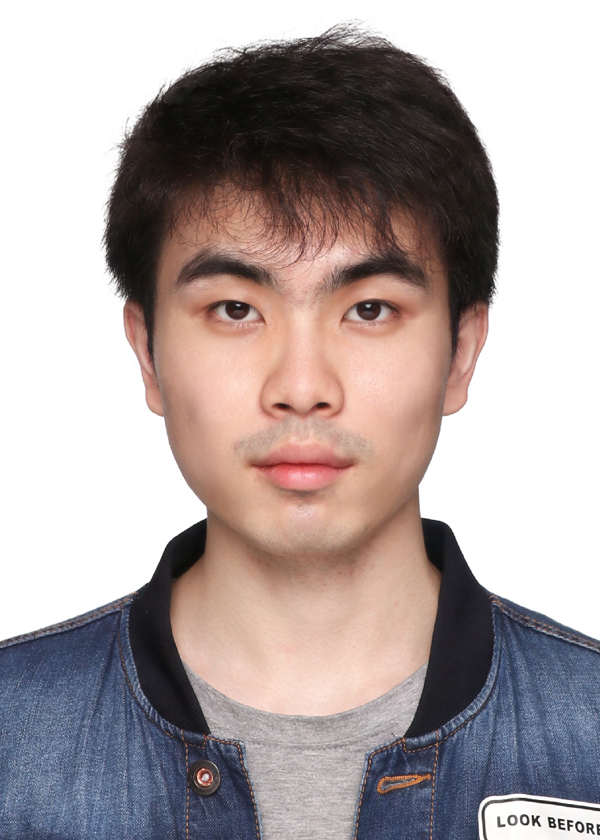}}]{Siyuan Shen}received the BS degree from Shanghaitech University, Shanghai, China, in 2019.
He is currently working toward the master’s
degree at ShanghaiTech University, Shanghai,
China. His research interests include non-line-of-sight imaging, computer vision and computational imaging.

\end{IEEEbiography}
\vspace{-10mm}
\begin{IEEEbiography}[{\includegraphics[width=1in,height=1.25in,clip,keepaspectratio]{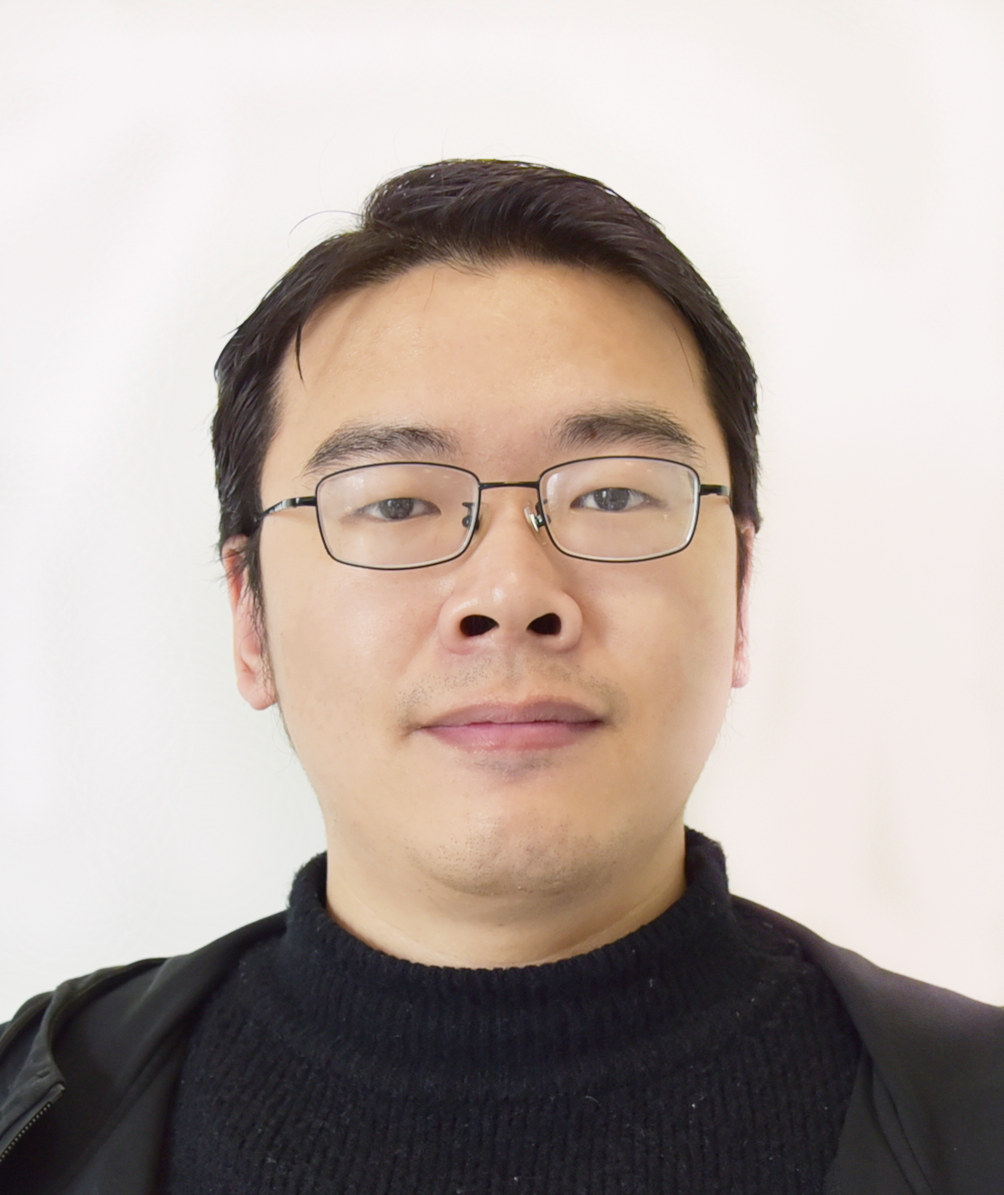}}]{Zi Wang} received the BS degree from the Beihang University, Beijing, China, in 2019.
He is currently working toward the master’s
degree at ShanghaiTech University, Shanghai,
China. His research interests include non-line-of-sight imaging and computational imaging.
\end{IEEEbiography}
\vspace{-10mm}
\begin{IEEEbiography}[{\includegraphics[width=1in,height=1.25in,clip,keepaspectratio]{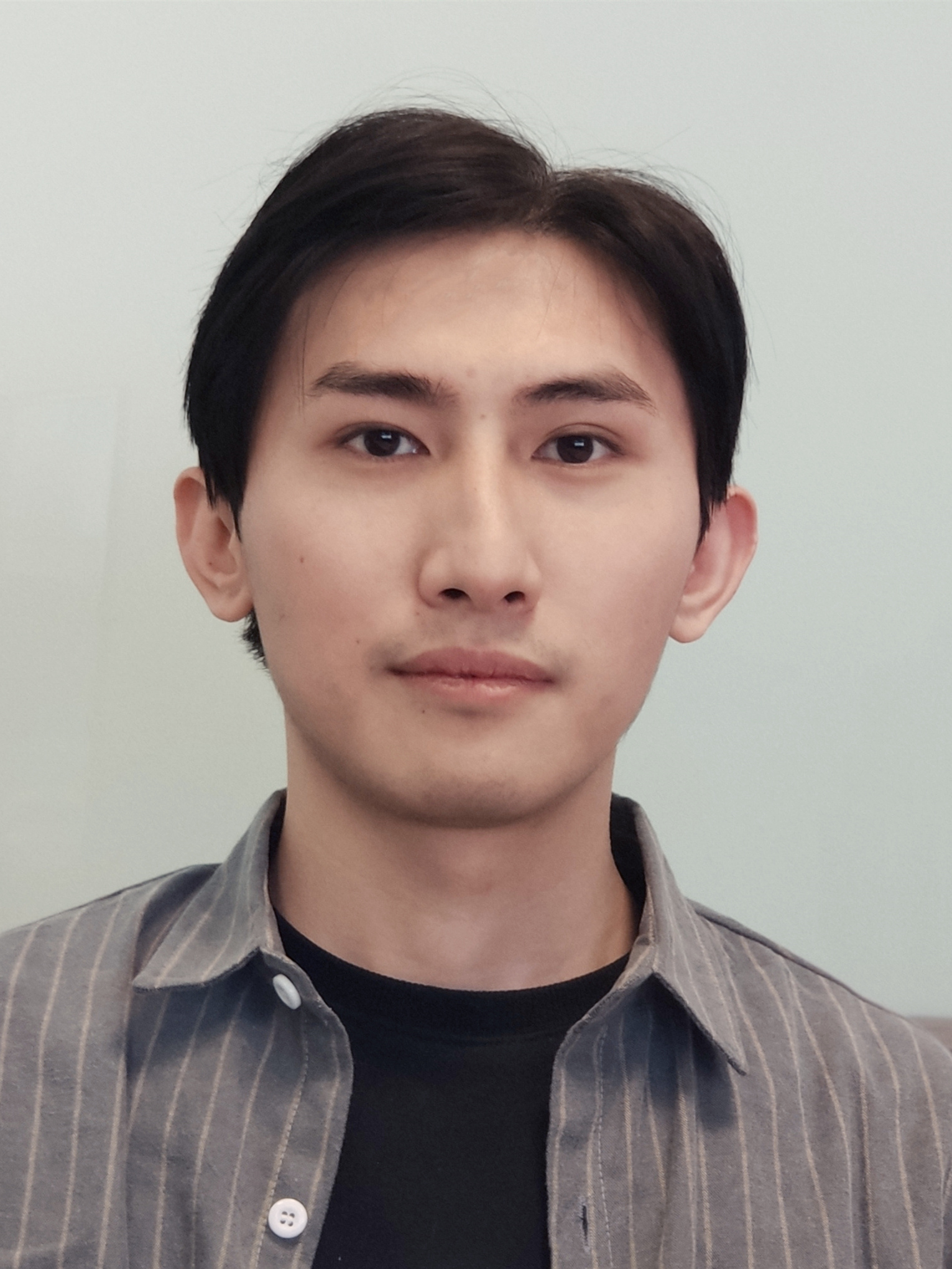}}]{Ping Liu}
received the BS degree from Central South University, Changsha, China, in 2020. He is working toward the master's degree in computer vision at ShanghaiTech University, Shanghai, China. His research interests include mainly in computational imaging and computer vision, especially on non-line-of-sight imaging.

\end{IEEEbiography}
\vspace{-8mm}
\begin{IEEEbiography}[{\includegraphics[width=1in,height=1.25in,clip,keepaspectratio]{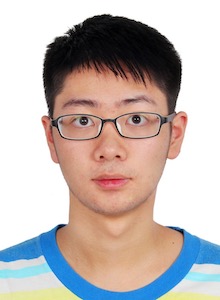}}]{Zhengqing Pan}
received the BS degree from Shanghaitech University, Shanghai, China, in 2019.
He is currently working toward the master’s
degree at ShanghaiTech University, Shanghai,
China. His research interests include single photon imaging and non-line-of-sight imaging.
\end{IEEEbiography}
\vspace{-8mm}
\begin{IEEEbiography}[{\includegraphics[width=1in,height=1.25in,clip,keepaspectratio]{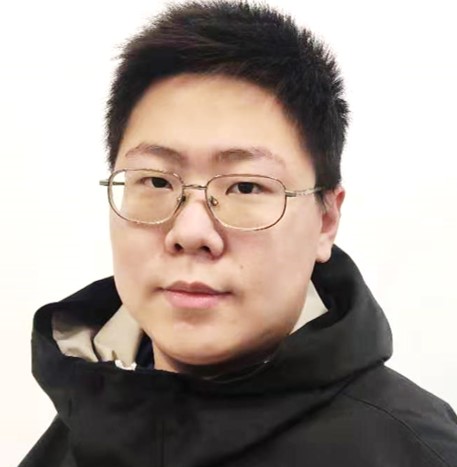}}]{Ruiqian Li}
is currently working toward the BS
degree at ShanghaiTech University, Shanghai,
China. His research interests include single photon imaging, noise model and denoising algorithm.
\end{IEEEbiography}
\vspace{-8mm}
\begin{IEEEbiography}[{\includegraphics[width=1in,height=1.25in,clip,keepaspectratio]{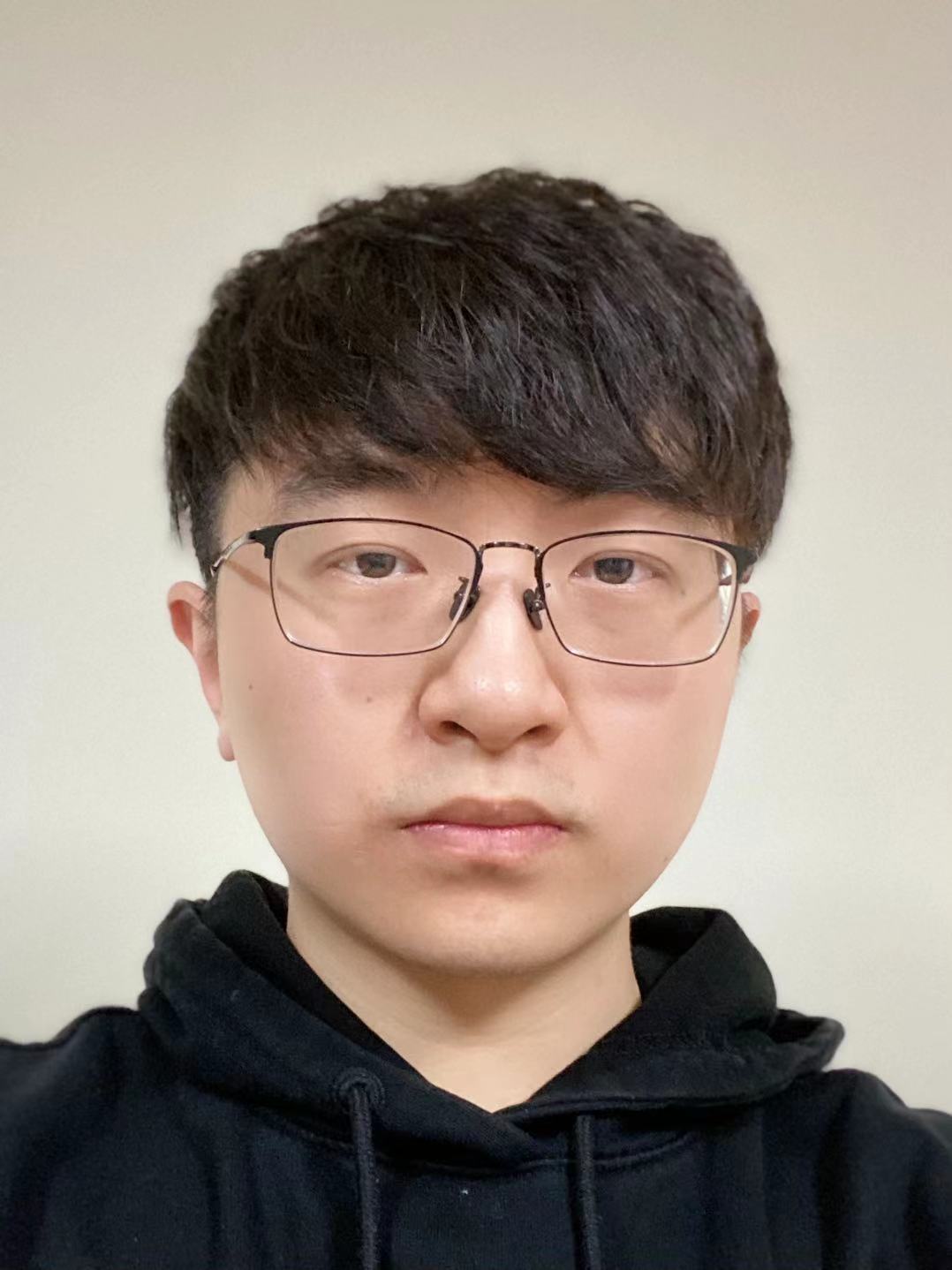}}]{Tian Gao}
He is currently working toward the BS
degree at ShanghaiTech University, Shanghai,
China. His research interests include single photon imaging and computational imaging.
\end{IEEEbiography}
\vspace{-8mm}
\begin{IEEEbiography}[{\includegraphics[width=1in,height=1.25in,clip,keepaspectratio]{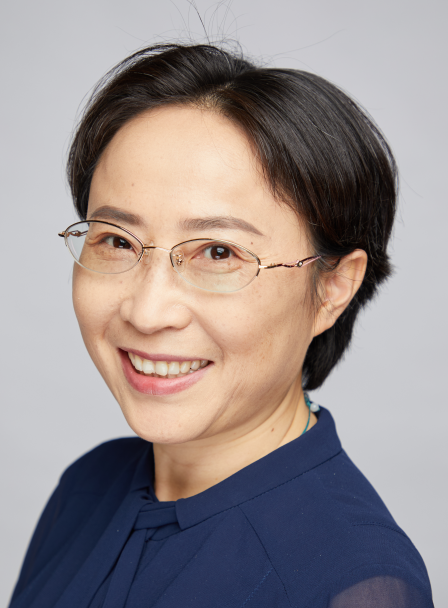}}]{Shiying Li} received her MS and PhD degrees in computer science from the Nara Institute of Science and Technology (NAIST), Nara, Japan, in 2004 and 2007. She worked as a post-doctoral research fellow in the Tohoku University from 2007 to 2008, Sendai, Japan. She was an associate professor with the Hunan University from 2009 to 2017, Changsha, Hunan, and has been an associate researcher with the School of Information Science and Technology, ShanghaiTech University, Shanghai, China. Her research interests include computational imaging, computer vision and graphics.  
\end{IEEEbiography}
\vspace{-8mm}
\begin{IEEEbiography}[{\includegraphics[width=1in,height=1.25in,clip,keepaspectratio]{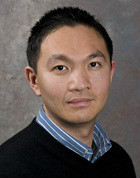}}]{Jingyi Yu} received BS from Caltech in 2000 and PhD from MIT in 2005. He is currently the Vice Provost at the ShanghaiTech University. Before joining ShanghaiTech, he was a full professor in the Department of Computer and Information Sciences at University of Delaware. His research interests span a range of topics in computer vision and computer graphics, especially on computational photography and non-conventional optics and camera designs. He is a recipient of the NSF CAREER Award and the AFOSR YIP Award, and has served as an area chair of many international conferences including CVPR, ICCV, ECCV, IJCAI and NeurIPS. He is currently a program chair of CVPR 2021 and will be a program chair of ICCV 2025. He has been an associate editor of the IEEE Transactions on Pattern Analysis and Machine Intelligence, the IEEE Transactions on Image Processing, and the Elsevier Computer Vision and Image Understanding. He is a fellow of IEEE. 
\end{IEEEbiography}



\fi

\begin{appendices}
\section{Formulating LCT via NeTF}
We show LCT can be formulated as a simplified NeTF model. We first rewrite the forward model Eqn.~\ref{tauSim} under triple integral with the Dirac delta function that correlates time of flight $t$ with distance $r$:
\begin{equation}
\begin{split}
    &\tau(x', y',t) = \\
    &\Gamma_{0} \underset{\Omega}{\iiint} \frac{\sin{\theta}}{r^{2}} \sigma(r, \theta,\phi) \rho(r, \theta,\phi) \delta(r - \frac{ct}{2}) \di{r}\di{\theta} \di{\phi}
\end{split}
\label{triple}
\end{equation}


\noindent where the integral domain $\Omega$ is defined under the spherical coordinates. Notice Eqn.~\ref{triple} is consistent with the light-cone transform (LCT) model~\cite{2018LCT}: we can rewrite it under the Cartesian coordinates where $\di{x}\di{y}\di{z} = {r^2 \sin{\theta}} {\di{r}\di{\theta} \di{\phi}}$ as:

\begin{equation}
\begin{split}
\tau&(x', y',t) = 2\Gamma_{0} \underset{\Omega}{\iiint} \frac{1}{r^4} \sigma(x,y,z) \rho(x,y,z,\theta,\phi) \cdot \\
& \delta(2\sqrt{(x - x')^2 + (y - y')^2 +z^2} - ct)\di{x}\di{y}\di{z}    
\end{split}
\label{after transform}
\end{equation}

If we further assume diffuse and isotropic albedo as $\rho_{\text{iso}}(x,y,z) = \sigma(x,y,z) \rho(x,y,z,\theta,\phi)$, Eqn.~\ref{after transform} degenerates to the LCT model (Eqn.~\ref{isoTau} with $g = 1$).



\section{Non-Confocal NeTF}

\begin{figure}[t!]
\centering
\includegraphics[width=1.0\columnwidth]{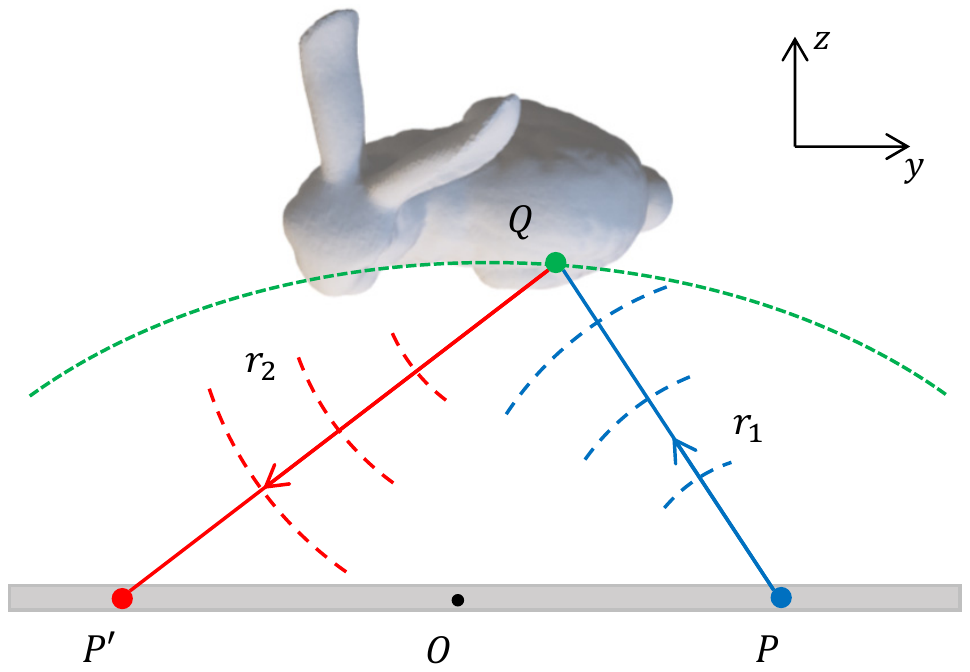}
\caption{Non-confocal NLOS imaging: The transient process from an illumination spot $P$ to an NLOS point $Q$ and from $Q$ to the detection spot $P'$ can be formulated under ellipsoidal coordinates.}
\label{nonconfocalfigure}
\end{figure}

Under the non-confocal setting, we set out to formulate the transient in terms of semi-ellipsoids with foci at the illumination and detection spots $P$ and $P'$ on the relay wall, as shown in Fig.~\ref{nonconfocalfigure}. Given a scene point $Q$, assume $r_1$ and $r_2$ correspond to the distance from $P$ to $Q$ and $Q$ to $P'$, respectively. Following the same derivations of Eqns.~\ref{EQ}, ~\ref{EQ'}, and~\ref{EP'} under the confocal setting, we first compute the energy (transient) received at $P'$ from the location $Q$ as:









\begin{equation}
E_{P'} = \frac{\Gamma}{r_2^{2}} \sigma(Q) \rho(Q,P,P') \exp{\left(-A\int_{\Upsilon}\sigma(s)\di{s}\right)} \di{\Omega}
\label{E'-PQP'.}
\end{equation}
where $\Gamma = Aar_0^2 E_p / \pi$. $\exp{\left(-A\int_{\Upsilon}\sigma(s)\di{s}\right)}$ corresponds to the attenuation coefficient along optical path $\Upsilon:P \to Q \to P'$ with length $r_1 + r_2 = ct$. 

To compute the complete transient received at $P'$ from $P$, recall $P'$ should be radiated by all points lying on a semi-ellipsoid $E$ with the foci $P, P'$, semi-major axis of length $\alpha = ct/2$, focal length $\gamma = |\overrightarrow{OP} - \overrightarrow{OP'}|$, and the eccentricity $e = \gamma / \alpha$. For simplicity, we can set up the coordinate system so that $P$ and $P'$ are symmetric about origin $O$ and $\overline{PP'}$ parallel to $y$-axis. We can thus compute transient as:
\begin{equation}
\begin{aligned}
\tau(P,P',t) = \underset{E}\iint E_{P'} \di{\Omega}
\label{EP'_nonconfocal}
\end{aligned}
\end{equation}

Since Eqn.~\ref{E'-PQP'.} is integrated on the semi-ellipsoid $E$ but under spherical coordinates centered at $P$, we need to rewrite $E$ under ellipsoidal coordinates with foci $P$ and $P'$. Specifically, we first represent the ellipsoid in terms of $r_1$ and $\theta$ as:

\begin{equation}
r_1 = \frac{\alpha(1 - e^{2})}{1 - e \cos{\theta}}
\label{spheroid-sph}
\end{equation}

Eqn.~\ref{EP'_nonconfocal} transforms to:

\begin{equation}
\begin{aligned}
&\tau(P,P',t) = \underset{\Omega}\iiint{E_{P'} \delta\left(r_1 - \frac{\alpha (1 - e^{2})}{1 - e\cos{\theta}}\right) \di{r_1} \di{\Omega}} \\
\label{E-PP'}
\end{aligned}
\end{equation}



Next, we transform spherical coordinates $(r_1, \theta, \phi)$ to ellipsoidal coordinates $(\mu,\nu,\varphi)$ as:
\begin{equation}
\begin{aligned}
r_1 \sin{\theta} \cos{\phi} &= \gamma \sinh{\mu} \sin{\nu} \cos{\varphi} \\
r_1 \sin{\theta} \sin{\phi} &= \gamma \sinh{\mu} \sin{\nu} \sin{\varphi} \\
r_1 \cos{\theta} &= \gamma \cosh{\mu} \cos{\nu}
\label{prolate-spheroidal}
\end{aligned}
\end{equation}

Recall that the Jacobian $J$ from the Cartesian to ellipsoidal coordinates are:
\begin{equation}
J =  \frac{\di{x} \di{y} \di{z}}{\di{\mu} \di{\nu} \di{\varphi}} = \gamma^{3} \sinh{\mu} \sin{\nu} (\sinh^{2}{\mu} + \sin^{2}{\nu})
\end{equation}

We can map between spherical coordinates to ellipsoidal coordinates via $J$ as:
\begin{equation}
\di{x} \di{y} \di{z} = r_1^2 \sin{\theta} \di{r_1} \di{\theta} \di{\phi} = r_1^2 \di{r_1} \di{\Omega}= J \di{\mu} \di{\nu} \di{\varphi}
\label{RtoE}
\end{equation}
Substituting Eqn.~\ref{RtoE} into Eqn.~\ref{E-PP'}, we obtain the transient under the ellipsoidal coordinate system as:

\begin{equation}
\begin{aligned}
\tau(P,P',t) = \underset{\Omega}\iiint \frac{1}{r_1^2} E_{P'} \delta\left(2\gamma\cosh{\mu}-ct\right) J \di{\mu} \di{\nu} \di{\varphi} 
\label{E'-PP'-}
\end{aligned}
\end{equation}

Notice that with a fixed $t$, we can find the corresponding $\mu$ for non-zero $\delta$ so that the triple integrate can be simplified to double integral in only $\nu$ and $\varphi$. In addition, if we further discard the attenuation term in $E_{P'}$, we can simplify the transient to: 
\begin{equation}
\begin{split}
    \tau(P,P',t) = \Gamma_{0} \underset{E}{\iint} \frac{J}{r_1^2 r_2^2}
    \sigma(\mu, \nu,\varphi) \rho(\mu, \nu,\varphi,P,P') \di{\nu} \di{\varphi}
\end{split}
\label{triple_non}
\end{equation}
where $\mu = arccosh(ct / 2\gamma)$.  In our non-confocal NeTF implementation (as in Sec.~\ref{Differential Rendering}), we set out to solve Eqn.~\ref{triple_non}. It is important to note though that a downside of discarding attenuation ignores occlusions.

\end{appendices}

\end{document}